\documentclass[12pt]{article}
\usepackage[utf8]{inputenc}
\usepackage{amsmath}
\usepackage{amsthm}
\usepackage{amsfonts}
\usepackage[numbers]{natbib}
\usepackage{bm}
\usepackage{thm-restate}

\usepackage{graphicx}
\usepackage{subcaption}
\usepackage{float}
\usepackage{authblk}

\usepackage{color}   
\usepackage{hyperref}
\hypersetup{
    linktoc=all,     
    linkcolor=blue,  
}

\usepackage{framed}
\usepackage{setspace}
\usepackage{listings}
\spacing{1.06}
\usepackage[margin=1in]{geometry}

\usepackage[ruled,vlined]{algorithm2e}

\theoremstyle{plain}

\theoremstyle{definition}

\theoremstyle{remark}

\usepackage{dsfont}

\begin{document}

\title{Lecture Notes from the NaijaCoder Summer Camp}

\author{
Daniel Alabi,
Joseph Ekpenyong,
Alida Monaco
}
\date{September 2024}

\maketitle

\begin{abstract}
 
The NaijaCoder in-person summer camps are intensive programs for
high school and pre-college students in Nigeria.
The programs are meant to provide free instruction on the basics of
algorithms and computer programming.
In 2024, the
camps were held in two locations within the country:
(i) the Federal Capital Territory (F.C.T.), Abuja; and (ii) Lagos state.
Both locations relied on the same set of notes for instructional purposes.
We are providing these notes in a publicly-available medium for both
students and teachers to review after the main in-person programs are over.

\begin{quote}
\quad\quad\quad\quad\quad``\textit{Learning is a journey that never ends.}''
\end{quote}

\end{abstract}

\clearpage

\tableofcontents
\clearpage

\section{Introduction}
The summer camps organized by NaijaCoder are programs aimed at
empowering young people in Nigeria with programming and computing
expertise skills. The initiative provides participants, typically
high-school students, with intensive training in various aspects of
programming, software development, and problem-solving. The program also
includes workshops and mentorship designed to equip attendees with the
skills needed to succeed in academia or in the tech
industry~\citep{AlabiAAAM24}.

The 2024 NaijaCoder program consisted of summer camps in Abuja and
Lagos, with collaboration from institutions like the University of
Lagos' AI \& Robotics Labs and Olumawu School in Abuja. Going forward,
NaijaCoder would also incorporate activities beyond the summer camp so
that students can keep improving their technical skills.

The lectures provided within this document are meant to enable groups of
students to: (i) spend more time to solve homework exercises they could
not solve during the program, (ii) learn about resources to keep
learning beyond the in-person bootcamps, (iii) serve as a reference as
they refresh their knowledge of basic concepts. The lectures are also
provided to teachers who can use the notes as a reference to keep
educating their students. (This document is not meant to be exhaustive
of all topics related to algorithms and computer programming.)

\paragraph{Textbook Reference}

In developing the lectures, the main reference textbook used was the
classic ``Introduction to Algorithms'\,' textbook
(CLRS)~\citep{CLRS}. Before proceeding through this
document, we encourage students to read Chapter 1 of
CLRS~\citep{CLRS}.

\paragraph{Further Motivation}

During the program, we had guest lectures on how to use algorithms and
computer programming to ``solve'\,' climate change. The goal of such
lectures was to provide motivation for why the students should learn how
to program even if they might not want a career as a software engineer.

\paragraph{Outline}

Most lecture notes begin with a reflection on topics or exercises from
the previous lecture. Then shortly after, the main topic for the lecture
is introduced.

\subsection{Introduction to the Basics of Python}

Python is a high-level, interpreted programming language known for its
readability, simplicity, and versatility. It was created by Guido van
Rossum and first released in 1991. Python's design philosophy emphasizes
code readability and simplicity, making it an ideal language for
beginners while still being powerful enough for advanced programming
tasks.

\subsubsection{Key Features of Python:}\label{key-features-of-python}

\begin{enumerate}
\def\labelenumi{\arabic{enumi}.}
\item
  \textbf{Readability}: Python uses clear, English-like syntax which
  makes it easier to learn and understand. This focus on readability
  reduces the cost of program maintenance.
\item
  \textbf{Interpreted Language}: Python code is executed line by line,
  which allows for easier debugging and quicker development cycles.
\item
  \textbf{Versatility}: Python can be used for a wide variety of
  applications, including web development, data analysis, artificial
  intelligence, scientific computing, automation, and more.
\item
  \textbf{Extensive Libraries and Frameworks}: Python has a vast
  ecosystem of libraries and frameworks, such as NumPy and Pandas for
  data analysis, TensorFlow and PyTorch for machine learning, Django and
  Flask for web development, and many others.
\item
  \textbf{Cross-Platform}: Python is cross-platform, meaning that Python
  programs can run on various operating systems, such as Windows, macOS,
  and Linux, without requiring significant changes.
\item
  \textbf{Community Support}: Python has a large and active community,
  which means there is a wealth of resources, documentation, and
  tutorials available for learners and developers.
\item
  \textbf{Object-Oriented}: Python supports object-oriented programming
  (OOP), which allows for organizing code into reusable classes and
  objects.
\end{enumerate}

\subsubsection{Common Uses of Python:}\label{common-uses-of-python}

\begin{itemize}
\item
  \textbf{Web Development}: Frameworks like Django and Flask make it
  easy to build web applications.
\item
  \textbf{Data Science and Machine Learning}: Libraries like Pandas,
  NumPy, Scikit-learn, and TensorFlow make Python the go-to language for
  data analysis and AI.
\item
  \textbf{Automation and Scripting}: Python is often used for automating
  repetitive tasks, such as file management or web scraping.
\item
  \textbf{Software Development}: Python can be used to develop software
  prototypes or full-fledged applications.
\end{itemize}

In Python, variables are used to store data values. Unlike some other
programming languages, Python does not require explicitly declaring the
data type of a variable when it is created. Instead, Python uses dynamic
typing, where variables can dynamically change their type based on the
value assigned to them.

\subsubsection{Variable Naming Rules}\label{variable-naming-rules}

When naming variables in Python, you need to follow these rules:

\begin{itemize}

\item
  Variable names must start with a letter (a-z, A-Z) or an underscore
  (\texttt{\_}). They cannot start with a number.
\item
  Variable names can only contain letters, numbers, and underscores
  (\texttt{\_}).
\item
  Variable names are case-sensitive (\texttt{age}, \texttt{Age}, and
  \texttt{AGE} are three different variables).
\item
  Python keywords (like \texttt{if}, \texttt{else}, \texttt{for}, etc.)
  cannot be used as variable names.
\end{itemize}

\subsubsection{Variable Assignment}\label{variable-assignment}

Variables in Python are assigned using the assignment operator
\texttt{=}. You can assign a value to a variable using this syntax:

\begin{verbatim}
variable_name = value
\end{verbatim}

\paragraph{Example 1: Assigning Values to
Variables}\label{example-1-assigning-values-to-variables}

\begin{verbatim}
# Assigning values to variables
name = "Alice"
age = 30
is_student = False
\end{verbatim}

\subsubsection{Variable Types}\label{variable-types}

Python supports various types of variables, including:

\begin{itemize}

\item
  \textbf{Integers (\texttt{int})}: Whole numbers without a decimal
  point.
\item
  \textbf{Floating-point numbers (\texttt{float})}: Numbers that have a
  decimal point or use exponential (e.g., \texttt{1.23},
  \texttt{3.45e-6}).
\item
  \textbf{Strings (\texttt{str})}: Textual data enclosed within quotes
  (\texttt{\textquotesingle{}single\textquotesingle{}} or
  \texttt{"double"}).
\item
  \textbf{Booleans (\texttt{bool})}: Represents truth values
  (\texttt{True} or \texttt{False}).
\item
  \textbf{Lists (\texttt{list})}: Ordered collections of items, mutable.
\item
  \textbf{Tuples (\texttt{tuple})}: Ordered collections of items,
  immutable.
\item
  \textbf{Dictionaries (\texttt{dict})}: Unordered collections of
  key-value pairs.
\end{itemize}

\subsubsection{Example: Variables of Different
Types}\label{example-variables-of-different-types}

\begin{verbatim}
# Variables of different types
age = 30              # Integer
average_grade = 85.5  # Float
student_name = "Bob"  # String
is_passing = True     # Boolean
\end{verbatim}

\subsubsection{Variable Reassignment}\label{variable-reassignment}

Variables in Python can be reassigned to different values of any type.
This is due to Python's dynamic typing nature.

\paragraph{Example 3: Reassigning
Variables}\label{example-3-reassigning-variables}

\begin{verbatim}
# Reassigning variables
age = 30
print(age)  # Output: 30

age = "thirty"
print(age)  # Output: thirty
\end{verbatim}

\subsubsection{Printing Variables}\label{printing-variables}

You can display the value of a variable using the \texttt{print()}
function.

\paragraph{Example 4: Printing
Variables}\label{example-4-printing-variables}

\begin{verbatim}
# Printing variables
name = "Alice"
print("Hello,", name)  # Output: Hello, Alice
\end{verbatim}

Variables in Python are essential for storing and manipulating data in
your programs. They are dynamically typed and flexible, allowing you to
work with different types of data seamlessly. Understanding how to
declare, assign, and use variables effectively is fundamental to writing
Python code.

In Python, comments are used to annotate code for various purposes such
as explaining functionality, providing documentation, or temporarily
disabling code. Comments are ignored by the Python interpreter during
execution and are purely for human readers. They help improve code
readability and understanding, especially for yourself and other
developers who may work on the code later.

\subsubsection{Types of Comments in
Python}\label{types-of-comments-in-python}

Python supports two types of comments:

\paragraph{1. Single-line Comments}\label{single-line-comments}

Single-line comments begin with the \texttt{\#} character and continue
until the end of the line. They are typically used for short comments on
a single line of code.

\subparagraph{Example of Single-line
Comment:}\label{example-of-single-line-comment}

\begin{verbatim}
# This is a single-line comment
x = 5  # Assigning value 5 to variable x
\end{verbatim}

\paragraph{2. Multi-line Comments
(Docstrings)}\label{multi-line-comments-docstrings}

Multi-line comments, also known as docstrings, are used for documenting
modules, classes, functions, or methods. They are enclosed in triple
quotes (\texttt{"""\ """} or
\texttt{\textquotesingle{}\textquotesingle{}\textquotesingle{}\ \textquotesingle{}\textquotesingle{}\textquotesingle{}})
and can span multiple lines. While docstrings are primarily used for
documentation, they can also be used as multi-line comments.

\subparagraph{Example of Multi-line Comment
(Docstring):}\label{example-of-multi-line-comment-docstring}

\begin{verbatim}
"""
This is a multi-line comment (docstring).
It can span multiple lines and is used for documentation.
"""
\end{verbatim}

\subsubsection{Best Practices for Using
Comments}\label{best-practices-for-using-comments}

\begin{itemize}

\item
  \textbf{Use Comments Sparingly}: Write code that is self-explanatory
  where possible. Use comments to clarify complex logic or important
  details.
\item
  \textbf{Be Clear and Concise}: Write clear comments that explain the
  intent or purpose of the code without stating the obvious.
\item
  \textbf{Update Comments}: Keep comments up-to-date with the code. If
  you modify the code's behavior, ensure the corresponding comments are
  also updated.
\item
  \textbf{Avoid Redundant Comments}: Avoid comments that simply repeat
  what the code is doing. Comments should provide additional context or
  explain why something is done a certain way.
\item
  \textbf{Use Docstrings for Documentation}: Use docstrings to document
  classes, functions, and methods according to Python's documentation
  conventions (PEP 257).
\end{itemize}

\subsubsection{Example of Good Commenting
Practices}\label{example-of-good-commenting-practices}

\begin{verbatim}
# Calculate the area of a rectangle
length = 5
width = 3
area = length * width  # Formula: length * width
print(f"The area of the rectangle is {area}")
\end{verbatim}

\subsubsection{When to Use Comments}\label{when-to-use-comments}

\begin{itemize}

\item
  \textbf{Explanation of Code}: Explain complex algorithms or logic that
  might not be immediately clear.
\item
  \textbf{TODO Comments}: Mark places where you intend to add more code
  or improvements in the future
  (\texttt{\#\ TODO:\ Implement\ error\ handling}).
\item
  \textbf{Debugging}: Temporarily comment out code for debugging
  purposes (\texttt{\#\ print(some\_variable)}).
\item
  \textbf{Documentation}: Use docstrings to document modules, classes,
  functions, and methods for generating documentation using tools like
  Sphinx.
\end{itemize}

Comments in Python are valuable for enhancing code readability,
explaining functionality, and providing documentation. By using comments
effectively and following best practices, you can create code that is
easier to understand, maintain, and collaborate on with other
developers.

\subsection{Exercises}\label{exercises}

Go through the following exercises in CLRS~\citep{CLRS}:

\begin{enumerate}
\item Exercise 1-1. (Use the python time module to solve the exercise.)
\item Exercises 1.1-1 up to 1.1-5. 
\item Exercises 1.2-1 up to 1.2-3.
\item Other than climate change, can you name 10 other areas that
computer programming can be applied to?
\end{enumerate}

\clearpage

\section{Types in Python}
\subsubsection{Reflection from Last
Day:}\label{reflection-from-last-day}

\begin{itemize}

\item
  Answer questions about Chapter 1 of CLRS.
\item
  Finish Chapter 1 exercises (1.1-1 up to 1.1-5. 1.2-1 up to 1.2-3. 1-1)
  of CLRS.
\end{itemize}

\subsubsection{\texorpdfstring{Integers
(\texttt{int}):}{Integers (int):}}\label{integers-int}

Integers are whole numbers, without a fractional component.

\begin{verbatim}
# Positive integer
a = 10

# Negative integer
b = -5

# Zero
c = 0
\end{verbatim}

\subsubsection{\texorpdfstring{Floating-Point Numbers
(\texttt{float}):}{Floating-Point Numbers (float):}}\label{floating-point-numbers-float}

Floating-point numbers have a decimal point and can represent fractional
values.

\begin{verbatim}
# Positive float
d = 10.5

# Negative float
e = -3.14

# Zero as a float
f = 0.0

# Scientific notation
g = 1.23e4  # This is equal to 12300.0
h = 1.23e-4  # This is equal to 0.000123
\end{verbatim}

\subsubsection{\texorpdfstring{Complex Numbers
(\texttt{complex}):}{Complex Numbers (complex):}}\label{complex-numbers-complex}

Complex numbers have a real and an imaginary part, represented as
\texttt{a\ +\ bj}, where \texttt{a} is the real part and \texttt{b} is
the imaginary part.

\begin{verbatim}
# Complex number with real and imaginary parts
i = 2 + 3j

# Complex number with only a real part
j = 5 + 0j

# Complex number with only an imaginary part
k = 0 + 7j
\end{verbatim}

Try these on your computer!

In Python, strings are sequences of characters enclosed in either single
quotes (\texttt{\textquotesingle{}}), double quotes (\texttt{"}), triple
single quotes
(\texttt{\textquotesingle{}\textquotesingle{}\textquotesingle{}}), or
triple double quotes (\texttt{"""}). Here are some examples
demonstrating the different ways to work with strings:

\subsubsection{Basic String Creation}\label{basic-string-creation}

\begin{verbatim}
# Single quotes
a = 'Hello, World!'
print(a)  # Output: Hello, World!

# Double quotes
b = "Hello, World!"
print(b)  # Output: Hello, World!

# Triple single quotes (for multi-line strings)
c = '''Hello,
World!'''
print(c)
# Output:
# Hello,
# World!

# Triple double quotes (for multi-line strings)
d = """Hello,
World!"""
print(d)
# Output:
# Hello,
# World!
\end{verbatim}

\subsubsection{String Concatenation}\label{string-concatenation}

\begin{verbatim}
# Using the + operator
e = "Hello" + ", " + "World!"
print(e)  # Output: Hello, World!

# Using f-strings (formatted string literals)
name = "Alice"
f = f"Hello, {name}!"
print(f)  # Output: Hello, Alice!

# Using the join method
g = " ".join(["Hello", "World!"])
print(g)  # Output: Hello World!
\end{verbatim}

\subsubsection{String Repetition}\label{string-repetition}

\begin{verbatim}
# Repeating a string multiple times
h = "Ha" * 3
print(h)  # Output: HaHaHa
\end{verbatim}

\subsubsection{String Indexing and
Slicing}\label{string-indexing-and-slicing}

\begin{verbatim}
# Indexing (accessing a single character)
i = "Hello"
print(i[0])  # Output: H
print(i[-1])  # Output: o

# Slicing (accessing a substring)
j = "Hello, World!"
print(j[0:5])  # Output: Hello
print(j[7:])  # Output: World!
print(j[:5])  # Output: Hello
print(j[::2])  # Output: Hlo ol!
\end{verbatim}

\subsubsection{String Methods}\label{string-methods}

\begin{verbatim}
# Convert to uppercase
k = "hello".upper()
print(k)  # Output: HELLO

# Convert to lowercase
l = "HELLO".lower()
print(l)  # Output: hello

# Strip whitespace
m = "  Hello  ".strip()
print(m)  # Output: Hello

# Replace substrings
n = "Hello, World!".replace("World", "Python")
print(n)  # Output: Hello, Python!

# Split into a list
o = "Hello, World!".split(", ")
print(o)  # Output: ['Hello', 'World!']

# Find a substring
p = "Hello, World!".find("World")
print(p)  # Output: 7 (index of the first occurrence of "World")
\end{verbatim}

\subsubsection{String Formatting}\label{string-formatting}

\begin{verbatim}
# Using the format method
q = "Hello, {}!".format("Alice")
print(q)  # Output: Hello, Alice!

# Using f-strings (formatted string literals)
r = f"Hello, {name}!"
print(r)  # Output: Hello, Alice!

# Using % operator
s = "Hello, %s!" % "Alice"
print(s)  # Output: Hello, Alice!
\end{verbatim}

\subsubsection{Multiline Strings with Line
Breaks}\label{multiline-strings-with-line-breaks}

\begin{verbatim}
# Using triple quotes
t = """This is a string
that spans multiple
lines."""
print(t)
# Output:
# This is a string
# that spans multiple
# lines.
\end{verbatim}

These examples cover the basics of string creation, manipulation, and
formatting in Python.

Here are some examples demonstrating boolean types (\texttt{bool}) in
Python, including their creation, usage in logical operations, and
conversion from other types:

\subsubsection{Basic Boolean Values}\label{basic-boolean-values}

\begin{verbatim}
# Boolean True
a = True
print(a)  # Output: True

# Boolean False
b = False
print(b)  # Output: False
\end{verbatim}

\subsubsection{Boolean from Comparisons}\label{boolean-from-comparisons}

\begin{verbatim}
# Comparison resulting in True
c = 5 > 3
print(c)  # Output: True

# Comparison resulting in False
d = 5 < 3
print(d)  # Output: False
\end{verbatim}

\subsubsection{Boolean Operations}\label{boolean-operations}

\begin{verbatim}
# Logical AND
e = True and False
print(e)  # Output: False

# Logical OR
f = True or False
print(f)  # Output: True

# Logical NOT
g = not True
print(g)  # Output: False
\end{verbatim}

\subsubsection{Using Boolean in Conditional
Statements}\label{using-boolean-in-conditional-statements}

\begin{verbatim}
# If statement
p = True
if p:
    print("This is True")  # Output: This is True

# If-else statement
q = False
if q:
    print("This is True")
else:
    print("This is False")  # Output: This is False
\end{verbatim}

\subsubsection{Boolean Methods and
Properties}\label{boolean-methods-and-properties}

\begin{verbatim}
# Checking if a boolean is True
r = True
if r is True:
    print("r is True")  # Output: r is True

# Checking if a boolean is False
s = False
if s is False:
    print("s is False")  # Output: s is False

# Using booleans in expressions
t = 10 > 5 and 5 < 20
print(t)  # Output: True

u = not (5 == 5)
print(u)  # Output: False
\end{verbatim}

These examples demonstrate the basic usage of boolean types, including
their values, operations, conversions, and use in conditional
statements.

In Python, \texttt{if} statements are used for conditional execution of
code based on evaluating expressions. The \texttt{if} statement allows
you to execute a block of code if a specified condition is true.
Optionally, you can also include \texttt{else} and \texttt{elif} (short
for ``else if'') statements to specify alternative blocks of code to
execute when the \texttt{if} condition is false or when additional
conditions need to be checked.

\subsubsection{\texorpdfstring{Basic \texttt{if}
Statement}{Basic if Statement}}\label{basic-if-statement}

The basic syntax of the \texttt{if} statement in Python is as follows:

\begin{verbatim}
if condition:
    # Block of code to execute if condition is true
    statement(s)
\end{verbatim}

\begin{itemize}

\item
  \textbf{\texttt{condition}}: An expression that evaluates to
  \texttt{True} or \texttt{False}.
\item
  \textbf{\texttt{statement(s)}}: Code to be executed if the condition
  is \texttt{True}. Indentation (typically 4 spaces) is crucial in
  Python to indicate the scope of the code block.
\end{itemize}

\paragraph{\texorpdfstring{Example 1: Simple \texttt{if}
Statement}{Example 1: Simple if Statement}}\label{example-1-simple-if-statement}

\begin{verbatim}
# Example of a simple if statement
x = 10
if x > 5:
    print("x is greater than 5")
\end{verbatim}

Output:

\begin{verbatim}
x is greater than 5
\end{verbatim}

\subsubsection{\texorpdfstring{\texttt{if}\ldots{}\texttt{else}
Statement}{if\ldots else Statement}}\label{ifelse-statement}

The \texttt{else} statement is used to execute a block of code when the
\texttt{if} condition is \texttt{False}. The syntax is:

\begin{verbatim}
if condition:
    # Block of code to execute if condition is true
    statement(s)
else:
    # Block of code to execute if condition is false
    statement(s)
\end{verbatim}

\paragraph{\texorpdfstring{Example 2: \texttt{if}\ldots{}\texttt{else}
Statement}{Example 2: if\ldots else Statement}}\label{example-2-ifelse-statement}

\begin{verbatim}
# Example of if...else statement
x = 2
if x > 5:
    print("x is greater than 5")
else:
    print("x is not greater than 5")
\end{verbatim}

Output:

\begin{verbatim}
x is not greater than 5
\end{verbatim}

\subsubsection{\texorpdfstring{\texttt{if}\ldots{}\texttt{elif}\ldots{}\texttt{else}
Statement}{if\ldots elif\ldots else Statement}}\label{ifelifelse-statement}

The \texttt{elif} statement allows you to check multiple conditions. It
is short for ``else if''. You can have multiple \texttt{elif} blocks
followed by an optional \texttt{else} block. The syntax is:

\begin{verbatim}
if condition1:
    # Block of code to execute if condition1 is true
    statement(s)
elif condition2:
    # Block of code to execute if condition2 is true
    statement(s)
else:
    # Block of code to execute if all conditions are false
    statement(s)
\end{verbatim}

\paragraph{\texorpdfstring{Example 3:
\texttt{if}\ldots{}\texttt{elif}\ldots{}\texttt{else}
Statement}{Example 3: if\ldots elif\ldots else Statement}}\label{example-3-ifelifelse-statement}

\begin{verbatim}
# Example of if...elif...else statement
x = 10
if x > 15:
    print("x is greater than 15")
elif x > 5:
    print("x is greater than 5 but not greater than 15")
else:
    print("x is 5 or less")
\end{verbatim}

Output:

\begin{verbatim}
x is greater than 5 but not greater than 15
\end{verbatim}

\subsubsection{\texorpdfstring{Nested \texttt{if}
Statements}{Nested if Statements}}\label{nested-if-statements}

You can also nest \texttt{if} statements within each other to create
more complex conditional logic.

\paragraph{\texorpdfstring{Example 4: Nested \texttt{if}
Statements}{Example 4: Nested if Statements}}\label{example-4-nested-if-statements}

\begin{verbatim}
# Example of nested if statements
x = 10
if x > 5:
    if x > 15:
        print("x is greater than 15")
    else:
        print("x is greater than 5 but not greater than 15")
else:
    print("x is 5 or less")
\end{verbatim}

Output:

\begin{verbatim}
x is greater than 5 but not greater than 15
\end{verbatim}

\subsubsection{Conditional Expressions (Ternary
Operator)}\label{conditional-expressions-ternary-operator}

Python also supports a concise way to write conditional expressions
known as the ternary operator \texttt{if}\ldots{}\texttt{else}
expression.

\paragraph{Example 5: Ternary
Operator}\label{example-5-ternary-operator}

\begin{verbatim}
# Ternary operator example
x = 10
result = "x is greater than 5" if x > 5 else "x is 5 or less"
print(result)  # Output: x is greater than 5
\end{verbatim}

In Python, \texttt{NoneType} is the type of the \texttt{None} object,
which is used to represent the absence of a value or a null value.
Here's an overview of \texttt{NoneType} and a simple use:

\subsubsection{Assigning None to a
variable}\label{assigning-none-to-a-variable}

\begin{verbatim}
a = None
print(a)  # Output: None
\end{verbatim}

\subsubsection{Checking if a variable is
None}\label{checking-if-a-variable-is-none}

\begin{verbatim}
if a is None:
    print("a is None")  # Output: a is None
else:
    print("a is not None")
\end{verbatim}

In Python, arrays can be created using different data structures, each
with its own use cases and characteristics. Here are some common types
of arrays in Python:

\subsubsection{Lists}\label{lists}

Lists are the most versatile array type in Python. They can store
elements of different types and support various operations.

\begin{verbatim}
# Creating a list
a = [1, 2, 3, 4, 5]
print(a)  # Output: [1, 2, 3, 4, 5]

# Accessing elements
print(a[0])  # Output: 1
print(a[-1])  # Output: 5

# Modifying elements
a[2] = 10
print(a)  # Output: [1, 2, 10, 4, 5]

# Adding elements
a.append(6)
print(a)  # Output: [1, 2, 10, 4, 5, 6]

# Removing elements
a.remove(10)
print(a)  # Output: [1, 2, 4, 5, 6]

# Slicing
print(a[1:4])  # Output: [2, 4, 5]
\end{verbatim}

\subsubsection{Tuples}\label{tuples}

Tuples are similar to lists but are immutable, meaning their elements
cannot be changed after creation.

\begin{verbatim}
# Creating a tuple
b = (1, 2, 3, 4, 5)
print(b)  # Output: (1, 2, 3, 4, 5)

# Accessing elements
print(b[0])  # Output: 1
print(b[-1])  # Output: 5

# Tuples are immutable, so elements cannot be modified
# b[2] = 10  # This will raise a TypeError

# Slicing
print(b[1:4])  # Output: (2, 3, 4)
\end{verbatim}

\subsubsection{\texorpdfstring{Arrays (from the \texttt{array}
module)}{Arrays (from the array module)}}\label{arrays-from-the-array-module}

The \texttt{array} module provides a way to create arrays with a fixed
type.

\begin{verbatim}
import array

# Creating an integer array
c = array.array('i', [1, 2, 3, 4, 5])
print(c)  # Output: array('i', [1, 2, 3, 4, 5])

# Accessing elements
print(c[0])  # Output: 1
print(c[-1])  # Output: 5

# Modifying elements
c[2] = 10
print(c)  # Output: array('i', [1, 2, 10, 4, 5])

# Adding elements
c.append(6)
print(c)  # Output: array('i', [1, 2, 10, 4, 5, 6])

# Removing elements
c.remove(10)
print(c)  # Output: array('i', [1, 2, 4, 5, 6])

# Slicing
print(c[1:4])  # Output: array('i', [2, 4, 5])
\end{verbatim}

\subsubsection{Lists of Lists (2D
Arrays)}\label{lists-of-lists-2d-arrays}

Lists of lists can be used to create multi-dimensional arrays.

\begin{verbatim}
# Creating a 2D list
f = [
    [1, 2, 3],
    [4, 5, 6],
    [7, 8, 9]
]
print(f)
# Output:
# [[1, 2, 3],
#  [4, 5, 6],
#  [7, 8, 9]]

# Accessing elements
print(f[0][0])  # Output: 1
print(f[1][2])  # Output: 6

# Modifying elements
f[0][0] = 10
print(f)
# Output:
# [[10, 2, 3],
#  [4, 5, 6],
#  [7, 8, 9]]
\end{verbatim}

These examples cover various array types in Python, including lists,
tuples, arrays from the \texttt{array} module, NumPy arrays, and lists
of lists for 2D arrays.

In Python, a tuple is an ordered and immutable collection of elements.
It's similar to a list but with the key difference that tuples cannot be
modified once created. Tuples are defined using parentheses \texttt{()}
and can contain elements of any data type, including other tuples. They
are commonly used to group related data elements together.

\subsubsection{Creating Tuples}\label{creating-tuples}

Tuples are created by enclosing elements within parentheses \texttt{()}
separated by commas \texttt{,}.

\paragraph{Example 1: Creating a
Tuple}\label{example-1-creating-a-tuple}

\begin{verbatim}
# Creating a tuple
my_tuple = (1, 2, 3, 4, 5)
print(my_tuple)  # Output: (1, 2, 3, 4, 5)
\end{verbatim}

\paragraph{Example 2: Tuple with Mixed Data
Types}\label{example-2-tuple-with-mixed-data-types}

\begin{verbatim}
# Tuple with mixed data types
mixed_tuple = ('apple', 3, True, 2.5)
print(mixed_tuple)  # Output: ('apple', 3, True, 2.5)
\end{verbatim}

\paragraph{Example 3: Nested Tuples}\label{example-3-nested-tuples}

\begin{verbatim}
# Nested tuple
nested_tuple = ('hello', (1, 2, 3), ['a', 'b', 'c'])
print(nested_tuple)  # Output: ('hello', (1, 2, 3), ['a', 'b', 'c'])
\end{verbatim}

\subsubsection{Accessing Elements in
Tuples}\label{accessing-elements-in-tuples}

Elements in a tuple are accessed using indexing, similar to lists.
Indexing starts at 0 for the first element.

\paragraph{Example 4: Accessing Tuple
Elements}\label{example-4-accessing-tuple-elements}

\begin{verbatim}
# Accessing elements in a tuple
print(my_tuple[0])  # Output: 1
print(my_tuple[3])  # Output: 4
\end{verbatim}

\subsubsection{Tuple Methods}\label{tuple-methods}

Since tuples are immutable, they have a limited number of methods
compared to lists. Here are some common methods:

\begin{itemize}

\item
  \textbf{\texttt{len()}}: Returns the number of elements in the tuple.
\item
  \textbf{\texttt{count()}}: Returns the number of occurrences of a
  specified value.
\item
  \textbf{\texttt{index()}}: Returns the index of the first occurrence
  of a specified value.
\end{itemize}

\paragraph{Example 5: Using Tuple
Methods}\label{example-5-using-tuple-methods}

\begin{verbatim}
# Tuple methods
my_tuple = (1, 2, 2, 3, 4, 2)
print(len(my_tuple))    # Output: 6
print(my_tuple.count(2))    # Output: 3
print(my_tuple.index(3))    # Output: 3
\end{verbatim}

\subsubsection{Iterating Through Tuples}\label{iterating-through-tuples}

You can iterate through tuples using \texttt{for} loops, similar to
lists and other iterable objects.

\paragraph{Example 6: Iterating Through a
Tuple}\label{example-6-iterating-through-a-tuple}

\begin{verbatim}
# Iterating through a tuple
for item in my_tuple:
    print(item)

# Output:
# 1
# 2
# 2
# 3
# 4
# 2
\end{verbatim}

\subsubsection{Advantages of Tuples}\label{advantages-of-tuples}

\begin{itemize}

\item
  \textbf{Immutable}: Once created, tuples cannot be modified, making
  them useful for protecting data integrity.
\item
  \textbf{Faster Access}: Tuples are generally faster for accessing
  elements compared to lists.
\item
  \textbf{Used as Dictionary Keys}: Tuples can be used as keys in
  dictionaries because they are immutable.
\end{itemize}

\subsubsection{When to Use Tuples}\label{when-to-use-tuples}

\begin{itemize}

\item
  Use tuples when you have a collection of items that should not change,
  such as coordinates, settings, or configuration values.
\item
  Use tuples as dictionary keys when the key is a fixed set of values.
\end{itemize}

\subsubsection{Defining Functions}\label{defining-functions}

\begin{verbatim}
# Simple function definition
def greet(name):
    return f"Hello, {name}!"

# Calling the function
print(greet("Alice"))  # Output: Hello, Alice!
\end{verbatim}

\subsubsection{Functions as First-Class
Objects}\label{functions-as-first-class-objects}

\begin{verbatim}
# Assigning a function to a variable
def add(a, b):
    return a + b

sum_function = add

# Calling the function through the variable
print(sum_function(3, 4))  # Output: 7
\end{verbatim}

\subsubsection{Passing Functions as
Arguments}\label{passing-functions-as-arguments}

\begin{verbatim}
# Function that takes another function as an argument
def operate(func, x, y):
    return func(x, y)

# Using the function
result = operate(add, 5, 3)
print(result)  # Output: 8
\end{verbatim}

\subsubsection{Returning Functions from
Functions}\label{returning-functions-from-functions}

\begin{verbatim}
# Function that returns another function
def multiplier(factor):
    def multiply_by_factor(number):
        return number * factor
    return multiply_by_factor

# Using the returned function
double = multiplier(2)
print(double(5))  # Output: 10
\end{verbatim}

\subsubsection{Higher-Order Functions}\label{higher-order-functions}

\begin{verbatim}
# Using built-in higher-order functions
numbers = [1, 2, 3, 4, 5]

# Using map function
squared_numbers = map(lambda x: x**2, numbers)
print(list(squared_numbers))  # Output: [1, 4, 9, 16, 25]

# Using filter function
even_numbers = filter(lambda x: x % 2 == 0, numbers)
print(list(even_numbers))  # Output: [2, 4]
\end{verbatim}

\subsubsection{Functions with Default
Arguments}\label{functions-with-default-arguments}

\begin{verbatim}
# Function with default arguments
def greet(name, greeting="Hello"):
    return f"{greeting}, {name}!"

# Using the function
print(greet("Alice"))           # Output: Hello, Alice!
print(greet("Bob", "Hi"))       # Output: Hi, Bob!
\end{verbatim}

\subsubsection{Functions with Variable-Length
Arguments}\label{functions-with-variable-length-arguments}

\begin{verbatim}
# Function with variable-length arguments (*args and **kwargs)
def example(*args, **kwargs):
    print("Positional arguments:", args)
    print("Keyword arguments:", kwargs)

# Using the function
example(1, 2, 3, key1='value1', key2='value2')
# Output:
# Positional arguments: (1, 2, 3)
# Keyword arguments: {'key1': 'value1', 'key2': 'value2'}
\end{verbatim}

\subsubsection{Anonymous (Lambda)
Functions}\label{anonymous-lambda-functions}

\begin{verbatim}
# Defining and using a lambda function
multiply = lambda x, y: x * y

print(multiply(3, 4))  # Output: 12

# Using a lambda function in a higher-order function
numbers = [1, 2, 3, 4, 5]
squared_numbers = map(lambda x: x**2, numbers)
print(list(squared_numbers))  # Output: [1, 4, 9, 16, 25]
\end{verbatim}

\subsubsection{Closures}\label{closures}

\begin{verbatim}
# Example of a closure
def outer_function(outer_var):
    def inner_function(inner_var):
        return outer_var + inner_var
    return inner_function

# Using the closure
add_five = outer_function(5)
print(add_five(10))  # Output: 15
\end{verbatim}

\subsubsection{Using Functions with
Decorators}\label{using-functions-with-decorators}

\begin{verbatim}
# Example of a decorator
def my_decorator(func):
    def wrapper():
        print("Something is happening before the function is called.")
        func()
        print("Something is happening after the function is called.")
    return wrapper

@my_decorator
def say_hello():
    print("Hello!")

# Calling the decorated function
say_hello()
# Output:
# Something is happening before the function is called.
# Hello!
# Something is happening after the function is called.
\end{verbatim}

Here are examples of converting between different number types in
Python:

\subsubsection{Converting Other Types to
Boolean}\label{converting-other-types-to-boolean}

\begin{verbatim}
# Integer to boolean (non-zero is True, zero is False)
h = bool(1)
print(h)  # Output: True

i = bool(0)
print(i)  # Output: False

# Float to boolean (non-zero is True, zero is False)
j = bool(0.1)
print(j)  # Output: True

k = bool(0.0)
print(k)  # Output: False

# String to boolean (non-empty string is True, empty string is False)
l = bool("Hello")
print(l)  # Output: True

m = bool("")
print(m)  # Output: False

# List to boolean (non-empty list is True, empty list is False)
n = bool([1, 2, 3])
print(n)  # Output: True

o = bool([])
print(o)  # Output: False
\end{verbatim}

\subsubsection{Converting Integers to
Floats}\label{converting-integers-to-floats}

\begin{verbatim}
# Integer
a = 10

# Convert to float
b = float(a)
print(b)  # Output: 10.0
\end{verbatim}

\subsubsection{Converting Floats to
Integers}\label{converting-floats-to-integers}

\begin{verbatim}
# Float
c = 10.5

# Convert to integer (note: this will truncate the decimal part)
d = int(c)
print(d)  # Output: 10
\end{verbatim}

\subsubsection{Converting Integers to Complex
Numbers}\label{converting-integers-to-complex-numbers}

\begin{verbatim}
# Integer
e = 5

# Convert to complex
f = complex(e)
print(f)  # Output: (5+0j)
\end{verbatim}

\subsubsection{Converting Floats to Complex
Numbers}\label{converting-floats-to-complex-numbers}

\begin{verbatim}
# Float
g = 7.1

# Convert to complex
h = complex(g)
print(h)  # Output: (7.1+0j)
\end{verbatim}

\subsubsection{Converting Strings to
Integers}\label{converting-strings-to-integers}

\begin{verbatim}
# String
i = "123"

# Convert to integer
j = int(i)
print(j)  # Output: 123
\end{verbatim}

\subsubsection{Converting Strings to
Floats}\label{converting-strings-to-floats}

\begin{verbatim}
# String
k = "123.45"

# Convert to float
l = float(k)
print(l)  # Output: 123.45
\end{verbatim}

These examples cover defining functions, using them as first-class
objects, passing them as arguments, returning them from other functions,
and using higher-order functions, default arguments, variable-length
arguments, lambda functions, closures, and decorators.

In Python, the \texttt{ord} and \texttt{chr} functions are used to
convert between characters and their corresponding ASCII (or Unicode)
code points.

\subsubsection{\texorpdfstring{\texttt{ord}
Function}{ord Function}}\label{ord-function}

The \texttt{ord} function takes a single character as an argument and
returns its corresponding ASCII (or Unicode) code point.

\begin{verbatim}
# Using ord to get the ASCII value of a character
ascii_value = ord('A')
print(ascii_value)  # Output: 65

# Using ord with another character
ascii_value = ord('a')
print(ascii_value)  # Output: 97
\end{verbatim}

\subsubsection{\texorpdfstring{\texttt{chr}
Function}{chr Function}}\label{chr-function}

The \texttt{chr} function takes an ASCII (or Unicode) code point and
returns the corresponding character.

\begin{verbatim}
# Using chr to get the character for an ASCII value
char = chr(65)
print(char)  # Output: A

# Using chr with another ASCII value
char = chr(97)
print(char)  # Output: a

# Using chr with a Unicode code point
char = chr(128512)
print(char)  # Output: a smily face
\end{verbatim}

\subsubsection{\texorpdfstring{Examples of Using \texttt{ord} and
\texttt{chr}
Together}{Examples of Using ord and chr Together}}\label{examples-of-using-ord-and-chr-together}

You can use \texttt{ord} and \texttt{chr} together to encode and decode
characters.

\begin{verbatim}
# Convert a character to its ASCII value and back to a character
char = 'B'
ascii_value = ord(char)
print(ascii_value)  # Output: 66

char_back = chr(ascii_value)
print(char_back)  # Output: B
\end{verbatim}

These examples show how to use \texttt{ord} and \texttt{chr} to work
with characters and their corresponding ASCII (or Unicode) values in
Python.

Here are examples demonstrating the use of dictionaries and sets in
Python, showcasing their features and common operations.

\subsubsection{Dictionaries}\label{dictionaries}

Dictionaries in Python are mutable, unordered collections of key-value
pairs. They are widely used for mapping keys to values and are denoted
by curly braces \texttt{\{\}}.

\paragraph{Example 1: Creating a
Dictionary}\label{example-1-creating-a-dictionary}

\begin{verbatim}
# Creating a dictionary
person = {'name': 'Alice', 'age': 30, 'city': 'Wonderland'}
print(person)  # Output: {'name': 'Alice', 'age': 30, 'city': 'Wonderland'}
\end{verbatim}

\paragraph{Example 2: Accessing
Values}\label{example-2-accessing-values}

\begin{verbatim}
# Accessing values in a dictionary
print(person['name'])  # Output: Alice
print(person.get('age'))  # Output: 30
\end{verbatim}

\paragraph{Example 3: Modifying and Adding
Elements}\label{example-3-modifying-and-adding-elements}

\begin{verbatim}
# Modifying values in a dictionary
person['age'] = 31
print(person)  # Output: {'name': 'Alice', 'age': 31, 'city': 'Wonderland'}

# Adding a new key-value pair
person['job'] = 'Engineer'
print(person)  # Output: {'name': 'Alice', 'age': 31, 'city': 'Wonderland', 'job': 'Engineer'}
\end{verbatim}

\paragraph{Example 4: Iterating Through a
Dictionary}\label{example-4-iterating-through-a-dictionary}

\begin{verbatim}
# Iterating through a dictionary
for key, value in person.items():
    print(f"{key}: {value}")

# Output:
# name: Alice
# age: 31
# city: Wonderland
# job: Engineer
\end{verbatim}

\paragraph{Example 5: Dictionary
Comprehension}\label{example-5-dictionary-comprehension}

\begin{verbatim}
# Dictionary comprehension to create a new dictionary
numbers = [1, 2, 3, 4, 5]
squared_numbers = {num: num**2 for num in numbers}
print(squared_numbers)  # Output: {1: 1, 2: 4, 3: 9, 4: 16, 5: 25}
\end{verbatim}

\subsubsection{Sets}\label{sets}

Sets in Python are unordered collections of unique elements, denoted by
curly braces \texttt{\{\}}.

\paragraph{Example 1: Creating a Set}\label{example-1-creating-a-set}

\begin{verbatim}
# Creating a set
fruits = {'apple', 'banana', 'orange'}
print(fruits)  # Output: {'orange', 'apple', 'banana'}
\end{verbatim}

\paragraph{Example 2: Adding and Removing
Elements}\label{example-2-adding-and-removing-elements}

\begin{verbatim}
# Adding elements to a set
fruits.add('pear')
print(fruits)  # Output: {'orange', 'apple', 'pear', 'banana'}

# Removing elements from a set
fruits.remove('banana')
print(fruits)  # Output: {'orange', 'apple', 'pear'}
\end{verbatim}

\paragraph{Example 3: Set Operations}\label{example-3-set-operations}

\begin{verbatim}
# Set operations (union, intersection, difference)
set1 = {1, 2, 3, 4, 5}
set2 = {4, 5, 6, 7, 8}

# Union
union_set = set1 | set2
print(union_set)  # Output: {1, 2, 3, 4, 5, 6, 7, 8}

# Intersection
intersection_set = set1 & set2
print(intersection_set)  # Output: {4, 5}

# Difference
difference_set = set1 - set2
print(difference_set)  # Output: {1, 2, 3}
\end{verbatim}

\paragraph{Example 4: Iterating Through a
Set}\label{example-4-iterating-through-a-set}

\begin{verbatim}
# Iterating through a set
for fruit in fruits:
    print(fruit)

# Output:
# orange
# apple
# pear
\end{verbatim}

\paragraph{Example 5: Set
Comprehension}\label{example-5-set-comprehension}

\begin{verbatim}
# Set comprehension to create a new set
numbers = [1, 2, 3, 4, 5, 5, 4, 3, 2, 1]
unique_numbers = {num for num in numbers}
print(unique_numbers)  # Output: {1, 2, 3, 4, 5}
\end{verbatim}

\subsection{Exercises}\label{exercises}

\textbackslash end\{markdown\}

Solve the following exercises:

\begin{enumerate}
\item Convert ``12345'' into a python list [1, 2, 3, 4, 5] and vice-versa. Can you put the code inside a function?
\item Add up the ASCII values of the following strings: ``Nigeria'', ``USA'', ``Africa'', ``Botswana'', ``Traveler'', ``misc'', ``madam'', ``@tuv''.
\item Write a function to check if a number is even.
\item Write a function to check if a number is odd.
\item Write a function to check if a number ($n$) is divisible by another number ($t$).
\item Write a function to return True if and only if a number is a prime number.
\item Write a function to return the prime factors of a number. 
\item Write a function to sum all the prime numbers in a list.
\item Write a function to reverse a string and shift every character. e.g., ``abcdef'' $\Rightarrow$ ``gfedcb'', ``123abc'' $\Rightarrow$ ``dcb432''.
\end{enumerate}

\clearpage

\section{For Loops and Recursion}
\subsubsection{Reflection from Last
Day:}\label{reflection-from-last-day}

\begin{itemize}

\item
  Answer questions about types in Python
\item
  Discuss exercises from previous day
\end{itemize}

\subsection{For Loops}\label{for-loops}

In Python, \texttt{for} loops are used to iterate over a sequence of
items, such as elements in a list, characters in a string, or any
iterable object. Here's an introduction to \texttt{for} loops in Python,
including basic syntax, examples of usage, and common patterns.

\subsubsection{Basic Syntax}\label{basic-syntax}

The general syntax of a \texttt{for} loop in Python is straightforward:

\begin{verbatim}
for item in iterable:
    # Do something with each item
    print(item)
\end{verbatim}

\begin{itemize}

\item
  \textbf{\texttt{item}}: This variable represents the current item in
  the iteration. You can name it anything you like.
\item
  \textbf{\texttt{iterable}}: This is the collection of items over which
  the loop iterates. It can be a list, tuple, string, range, or any
  iterable object.
\end{itemize}

\subsubsection{Example: Iterating over a
List}\label{example-iterating-over-a-list}

\begin{verbatim}
# Iterating over a list of numbers
numbers = [1, 2, 3, 4, 5]
for num in numbers:
    print(num)
# Output:
# 1
# 2
# 3
# 4
# 5
\end{verbatim}

\subsubsection{Example: Iterating over a
String}\label{example-iterating-over-a-string}

\begin{verbatim}
# Iterating over a string
message = "Hello"
for char in message:
    print(char)
# Output:
# H
# e
# l
# l
# o
\end{verbatim}

\subsubsection{Example: Iterating over a
Range}\label{example-iterating-over-a-range}

\begin{verbatim}
# Iterating over a range of numbers
for i in range(1, 6):  # Range from 1 to 5 (inclusive)
    print(i)
# Output:
# 1
# 2
# 3
# 4
# 5
\end{verbatim}

\subsubsection{\texorpdfstring{\texttt{break} and \texttt{continue}
Statements}{break and continue Statements}}\label{break-and-continue-statements}

You can use \texttt{break} to exit the loop prematurely and
\texttt{continue} to skip the current iteration:

\begin{verbatim}
# Using break in a loop
numbers = [1, 2, 3, 4, 5]
for num in numbers:
    if num == 3:
        break
    print(num)
# Output:
# 1
# 2

# Using continue in a loop
numbers = [1, 2, 3, 4, 5]
for num in numbers:
    if num == 3:
        continue
    print(num)
# Output:
# 1
# 2
# 4
# 5
\end{verbatim}

\subsubsection{\texorpdfstring{Nested \texttt{for}
Loops}{Nested for Loops}}\label{nested-for-loops}

You can nest \texttt{for} loops to iterate over multiple sequences or
create patterns:

\begin{verbatim}
# Nested for loops
for i in range(1, 4):
    for j in range(1, 4):
        print(i, j)
# Output:
# 1 1
# 1 2
# 1 3
# 2 1
# 2 2
# 2 3
# 3 1
# 3 2
# 3 3
\end{verbatim}

\subsubsection{\texorpdfstring{Using \texttt{enumerate} for Index and
Value}{Using enumerate for Index and Value}}\label{using-enumerate-for-index-and-value}

You can use \texttt{enumerate} to iterate over an iterable and get both
the index and value:

\begin{verbatim}
# Using enumerate
colors = ['red', 'green', 'blue']
for index, color in enumerate(colors):
    print(index, color)
# Output:
# 0 red
# 1 green
# 2 blue
\end{verbatim}

\subsubsection{\texorpdfstring{Iterating Over Multiple Lists with
\texttt{zip}}{Iterating Over Multiple Lists with zip}}\label{iterating-over-multiple-lists-with-zip}

You can use \texttt{zip} to iterate over multiple lists simultaneously:

\begin{verbatim}
# Using zip to iterate over multiple lists
names = ['Alice', 'Bob', 'Charlie']
ages = [25, 30, 35]
for name, age in zip(names, ages):
    print(f"{name} is {age} years old")
# Output:
# Alice is 25 years old
# Bob is 30 years old
# Charlie is 35 years old
\end{verbatim}

\subsection{Recursion}\label{recursion}

Recursion is a programming technique where a function calls itself to
solve smaller instances of the same problem. It's particularly useful
for problems that can be divided into similar sub-problems. Here's an
introduction to recursion in Python, including basic concepts, examples,
and key considerations.

\subsubsection{Basic Concepts}\label{basic-concepts}

\begin{enumerate}
\def\labelenumi{\arabic{enumi}.}

\item
  \textbf{Base Case}: The condition under which the recursive function
  stops calling itself. Without a base case, the function would call
  itself indefinitely, leading to a stack overflow.
\item
  \textbf{Recursive Case}: The part of the function where it calls
  itself with modified arguments, working towards the base case.
\end{enumerate}

\subsubsection{Example: Factorial
Function}\label{example-factorial-function}

The factorial of a non-negative integer ( n ) is the product of all
positive integers less than or equal to ( n ). It's denoted as ( n! ).
The factorial function can be defined recursively as follows:

\begin{itemize}
\item \( 0! = 1 \) (base case)
\item \( n! = n \times (n-1)! \) for \( n > 0 \) (recursive case)
\end{itemize}

\begin{verbatim}
def factorial(n):
    # Base case: factorial of 0 is 1
    if n == 0:
        return 1
    # Recursive case: n * factorial of (n-1)
    else:
        return n * factorial(n - 1)

# Testing the factorial function
print(factorial(5))  # Output: 120
\end{verbatim}

\subsubsection{Example: Fibonacci
Sequence}\label{example-fibonacci-sequence}

The Fibonacci sequence is a series of numbers where each number is the
sum of the two preceding ones. It can be defined recursively as follows:

\begin{itemize}

\item
  ( F(0) = 0 ) (base case)
\item
  ( F(1) = 1 ) (base case)
\item
  ( F(n) = F(n-1) + F(n-2) ) for ( n \textgreater{} 1 ) (recursive case)
\end{itemize}

\begin{verbatim}
def fibonacci(n):
    # Base cases
    if n == 0:
        return 0
    elif n == 1:
        return 1
    # Recursive case
    else:
        return fibonacci(n - 1) + fibonacci(n - 2)

# Testing the fibonacci function
print(fibonacci(6))  # Output: 8
\end{verbatim}

\subsubsection{Key Considerations}\label{key-considerations}

\begin{enumerate}
\def\labelenumi{\arabic{enumi}.}

\item
  \textbf{Base Case}: Ensure that there is a base case to terminate the
  recursion. Without it, the function will recurse indefinitely and
  cause a stack overflow.
\item
  \textbf{Recursive Call}: Each recursive call should progress towards
  the base case to avoid infinite recursion.
\item
  \textbf{Performance}: Recursive solutions can be less efficient than
  iterative ones due to the overhead of multiple function calls. For
  instance, the naive recursive Fibonacci implementation has exponential
  time complexity. Consider using memoization or iterative solutions for
  performance-critical applications.
\end{enumerate}

\subsubsection{Example: Sum of a List}\label{example-sum-of-a-list}

Here's an example of summing the elements of a list using recursion:

\begin{verbatim}
def sum_list(numbers):
    # Base case: empty list
    if not numbers:
        return 0
    # Recursive case: sum of the first element and the sum of the rest of the list
    else:
        return numbers[0] + sum_list(numbers[1:])

# Testing the sum_list function
print(sum_list([1, 2, 3, 4, 5]))  # Output: 15
\end{verbatim}

\subsubsection{Recursion vs.~Iteration}\label{recursion-vs.-iteration}

\begin{itemize}

\item
  \textbf{Recursion} is often more elegant and easier to understand for
  problems that have a natural recursive structure, such as tree
  traversals.
\item
  \textbf{Iteration} can be more efficient in terms of memory and
  performance for problems that don't inherently benefit from recursion.
\end{itemize}

\subsubsection{Conclusion}\label{conclusion}

\texttt{for} loops are fundamental in Python for iterating over
sequences and collections of data. They provide a concise way to perform
repetitive tasks and are flexible enough to handle various types of data
structures and iterations. Understanding these basics is crucial for
effective programming in Python.

Recursion is a powerful tool for solving problems that can be broken
down into smaller, similar sub-problems. By understanding the base case
and recursive case, you can effectively use recursion to solve a wide
range of problems. However, always be mindful of performance
implications and consider iterative solutions when appropriate.

\subsection{Exercises}\label{exercises}

Here are some exercises to practice \texttt{for} loops and recursion in
Python. These exercises range from basic to more advanced, covering
different aspects of iteration and recursion techniques.

\subsubsection{\texorpdfstring{Exercises for \texttt{for}
Loops}{Exercises for for Loops}}\label{exercises-for-for-loops}

\begin{enumerate}
\def\labelenumi{\arabic{enumi}.}
\item
  \textbf{Sum of Numbers}: Write a function that calculates the sum of
  all numbers from 1 to \texttt{n} using a \texttt{for} loop.
\item
  \textbf{Even Numbers}: Write a function that prints all even numbers
  from 1 to \texttt{n} using a \texttt{for} loop.
\item
  \textbf{Factorial}: Write a function to compute the factorial of a
  number \texttt{n} using a \texttt{for} loop.
\item
  \textbf{Printing Patterns}: Write a program to print the following
  pattern using nested \texttt{for} loops:

\begin{verbatim}
*
* *
* * *
* * * *
* * * * *
\end{verbatim}
\item
  \textbf{Fibonacci Sequence}: Write a function to print the first
  \texttt{n} numbers in the Fibonacci sequence using a \texttt{for}
  loop.
\item
  \textbf{List Comprehension}: Convert a list of integers into their
  squares using list comprehension and print the result.
\item
  \textbf{Multiplication Table}: Write a function that prints the
  multiplication table (up to 10) using nested \texttt{for} loops.
\item
  \textbf{Prime Numbers}: Write a function to print all prime numbers up
  to \texttt{n} using a \texttt{for} loop and a helper function.
\item
  \textbf{Character Pyramid}: Write a function that prints a pyramid of
  characters up to a given height \texttt{n} using nested \texttt{for}
  loops.
\end{enumerate}

\subsubsection{Exercises for Recursion}\label{exercises-for-recursion}

\begin{enumerate}
\def\labelenumi{\arabic{enumi}.}
\item
  \textbf{Factorial Recursion}: Rewrite the factorial function using
  recursion.
\item
  \textbf{Fibonacci Recursion}: Write a recursive function to compute
  the Fibonacci sequence up to \texttt{n} numbers.
\item
  \textbf{Power Function}: Write a recursive function to calculate the
  power of a number \texttt{base} raised to an exponent \texttt{exp}.
\item
  \textbf{Sum of Digits}: Write a recursive function to calculate the
  sum of digits of a positive integer.
\item
  \textbf{Binary Search}: Implement the binary search algorithm
  recursively to find an element in a sorted list.
\item
  \textbf{Greatest Common Divisor (GCD)}: Write a recursive function to
  find the GCD of two numbers using Euclid's algorithm.
\item
  \textbf{Merge Sort}: Implement the merge sort algorithm (to be
  introduced at a subsequent lecture) using recursion to sort a list of
  integers.
\item
  \textbf{Towers of Hanoi}: Implement the Towers of Hanoi problem using
  recursion.
\item
  \textbf{Palindrome Check}: Write a recursive function to check if a
  given string is a palindrome.
\item
  \textbf{Print Paths in a Grid}: Write a recursive function to print
  all possible paths from the top-left corner to the bottom-right corner
  of a \texttt{m\ x\ n} grid.
\end{enumerate}

\clearpage

\section{More Methods of Iteration in Python}
\subsubsection{Reflection from Last
Day:}\label{reflection-from-last-day}

\begin{itemize}

\item
  Answer questions about for loops and recursion in Python
\item
  Discuss exercises from previous day
\end{itemize}

Iteration in Python refers to the process of repeatedly executing a
block of code, typically involving the traversal of elements in a
sequence or collection. Iteration is fundamental in programming for
tasks such as processing each element in a list, iterating through
characters in a string, or executing a block of code a specified number
of times. Python provides several mechanisms for iteration, primarily
through \texttt{for} loops, \texttt{while} loops, and comprehensions.

\subsubsection{\texorpdfstring{\texttt{for}
Loops}{for Loops}}\label{for-loops}

The \texttt{for} loop in Python is used to iterate over elements of a
sequence, such as lists, tuples, strings, dictionaries, and other
iterable objects. It follows this basic syntax:

\begin{verbatim}
for item in iterable:
    # Code block to execute for each item
    print(item)
\end{verbatim}

\begin{itemize}

\item
  \textbf{\texttt{item}}: This variable represents the current element
  in the iteration.
\item
  \textbf{\texttt{iterable}}: This is the sequence or collection of
  items over which the loop iterates.
\end{itemize}

\textbf{Examples of \texttt{for} Loops:}

\begin{enumerate}
\def\labelenumi{\arabic{enumi}.}
\item
  \textbf{Iterating over a List:}

\begin{verbatim}
numbers = [1, 2, 3, 4, 5]
for num in numbers:
    print(num)
\end{verbatim}

  Output:

\begin{verbatim}
1
2
3
4
5
\end{verbatim}
\item
  \textbf{Iterating over a String:}

\begin{verbatim}
message = "Hello"
for char in message:
    print(char)
\end{verbatim}

  Output:

\begin{verbatim}
H
e
l
l
o
\end{verbatim}
\end{enumerate}

\subsubsection{\texorpdfstring{\texttt{while}
Loops}{while Loops}}\label{while-loops}

The \texttt{while} loop in Python repeatedly executes a block of code as
long as a specified condition is \texttt{True}. It follows this basic
syntax:

\begin{verbatim}
while condition:
    # Code block to execute as long as condition is True
    statement(s)
\end{verbatim}

\begin{itemize}

\item
  \textbf{\texttt{condition}}: This is an expression evaluated before
  each iteration. If \texttt{True}, the loop continues; if
  \texttt{False}, the loop terminates.
\end{itemize}

\textbf{Example of \texttt{while} Loop:}

\begin{verbatim}
count = 0
while count < 5:
    print(count)
    count += 1
\end{verbatim}

Output:

\begin{verbatim}
0
1
2
3
4
\end{verbatim}

\subsubsection{Comprehensions}\label{comprehensions}

Python comprehensions provide a concise way to create sequences such as
lists, dictionaries, and sets. They are often used for generating
sequences iteratively in a single line of code.

\textbf{List Comprehension Example:}

\begin{verbatim}
# Creating a list of squares of numbers from 1 to 5
squares = [x**2 for x in range(1, 6)]
print(squares)  # Output: [1, 4, 9, 16, 25]
\end{verbatim}

\textbf{Dictionary Comprehension Example:}

\begin{verbatim}
# Creating a dictionary from lists of keys and values
keys = ['a', 'b', 'c']
values = [1, 2, 3]
dictionary = {k: v for k, v in zip(keys, values)}
print(dictionary)  # Output: {'a': 1, 'b': 2, 'c': 3}
\end{verbatim}

\subsubsection{Key Concepts}\label{key-concepts}

\begin{itemize}

\item
  \textbf{Iterables}: Objects capable of returning their members one at
  a time, such as lists, tuples, strings, and dictionaries.
\item
  \textbf{Iterators}: Objects used to iterate over iterables, typically
  generated by \texttt{iter()} function.
\item
  \textbf{Generator Expressions}: Lazy evaluated iterables, often used
  to create iterators using
  \texttt{(expression\ for\ item\ in\ iterable)} syntax.
\end{itemize}

In Python, the \texttt{range()} function is used to generate a sequence
of numbers. It's commonly used in \texttt{for} loops to iterate over a
sequence of numbers rather than using lists of numbers directly. The
\texttt{range()} function returns a sequence of numbers based on the
arguments provided, which can be one, two, or three integers.

\subsubsection{Basic Syntax}\label{basic-syntax}

The basic syntax of the \texttt{range()} function is:

\begin{verbatim}
range(start, stop, step)
\end{verbatim}

\begin{itemize}

\item
  \textbf{\texttt{start}}: Optional. The starting value of the sequence.
  Default is \texttt{0}.
\item
  \textbf{\texttt{stop}}: Required. The ending value of the sequence
  (exclusive).
\item
  \textbf{\texttt{step}}: Optional. The step or increment between each
  number in the sequence. Default is \texttt{1}.
\end{itemize}

\subsubsection{Examples}\label{examples}

\paragraph{\texorpdfstring{Example 1: Using \texttt{range()} with one
argument}{Example 1: Using range() with one argument}}\label{example-1-using-range-with-one-argument}

When only one argument is provided, \texttt{range(stop)}, the sequence
starts from \texttt{0} and ends at \texttt{stop\ -\ 1}.

\begin{verbatim}
# Example using range(stop)
for i in range(5):
    print(i)
\end{verbatim}

Output:

\begin{verbatim}
0
1
2
3
4
\end{verbatim}

\paragraph{\texorpdfstring{Example 2: Using \texttt{range()} with two
arguments}{Example 2: Using range() with two arguments}}\label{example-2-using-range-with-two-arguments}

When two arguments are provided, \texttt{range(start,\ stop)}, the
sequence starts from \texttt{start} and ends at \texttt{stop\ -\ 1}.

\begin{verbatim}
# Example using range(start, stop)
for i in range(2, 7):
    print(i)
\end{verbatim}

Output:

\begin{verbatim}
2
3
4
5
6
\end{verbatim}

\paragraph{\texorpdfstring{Example 3: Using \texttt{range()} with three
arguments}{Example 3: Using range() with three arguments}}\label{example-3-using-range-with-three-arguments}

When three arguments are provided, \texttt{range(start,\ stop,\ step)},
the sequence starts from \texttt{start}, ends at \texttt{stop\ -\ 1},
and increments by \texttt{step}.

\begin{verbatim}
# Example using range(start, stop, step)
for i in range(1, 10, 2):
    print(i)
\end{verbatim}

Output:

\begin{verbatim}
1
3
5
7
9
\end{verbatim}

\subsubsection{\texorpdfstring{Generating Lists with
\texttt{range()}}{Generating Lists with range()}}\label{generating-lists-with-range}

You can convert the \texttt{range()} object into a list using
\texttt{list()} to see the actual sequence of numbers generated:

\begin{verbatim}
# Converting range to list
numbers = list(range(1, 6))
print(numbers)  # Output: [1, 2, 3, 4, 5]
\end{verbatim}

\subsubsection{\texorpdfstring{Negative Step with
\texttt{range()}}{Negative Step with range()}}\label{negative-step-with-range}

You can use a negative step to generate numbers in reverse order:

\begin{verbatim}
# Using negative step
for i in range(10, 0, -2):
    print(i)
\end{verbatim}

Output:

\begin{verbatim}
10
8
6
4
2
\end{verbatim}

\subsubsection{\texorpdfstring{Usage in \texttt{for}
Loops}{Usage in for Loops}}\label{usage-in-for-loops}

The \texttt{range()} function is often used in \texttt{for} loops to
iterate over a sequence of numbers:

\begin{verbatim}
# Iterating over a range of numbers
for i in range(1, 6):
    print(i)
\end{verbatim}

Output:

\begin{verbatim}
1
2
3
4
5
\end{verbatim}

\subsubsection{Performance
Considerations}\label{performance-considerations}

\begin{itemize}
\item
  \textbf{Memory Efficient}: Unlike creating a list with a sequence of
  numbers, \texttt{range()} generates numbers on the fly, making it
  memory efficient for large ranges.
\item
  \textbf{Immutable}: \texttt{range()} objects are immutable, meaning
  you cannot modify them directly.
\end{itemize}

\subsection{Exercises}\label{exercises}

\subsubsection{\texorpdfstring{Exercises on \texttt{for} and
\texttt{while}
Loops}{Exercises on for and while Loops}}\label{exercises-on-for-and-while-loops}

\begin{enumerate}
\def\labelenumi{\arabic{enumi}.}
\item
  \textbf{Sum of Numbers}: Write a function that calculates the sum of
  all numbers from 1 to \texttt{n} using \texttt{for}, \texttt{while}
  loops.
\item
  \textbf{Even Numbers}: Write a function that prints all even numbers
  from 1 to \texttt{n} using \texttt{for}, \texttt{while} loops.
\item
  \textbf{Factorial}: Write a function to compute the factorial of a
  number \texttt{n} using \texttt{for}, \texttt{while} loops.
\item
  \textbf{Printing Patterns}: Write a program to print the following
  pattern using nested using \texttt{for}, \texttt{while} loops.

\begin{verbatim}
*
* *
* * *
* * * *
* * * * *
\end{verbatim}
\item
  \textbf{Fibonacci Sequence}: Write a function to print the first
  \texttt{n} numbers in the Fibonacci sequence using \texttt{for},
  \texttt{while} loops.
\item
  \textbf{Multiplication Table}: Write a function that prints the
  multiplication table (up to 10) using nested \texttt{for},
  \texttt{while} loops.
\item
  \textbf{Prime Numbers}: Write a function to print all prime numbers up
  to \texttt{n} using a \texttt{for} loop and a helper function.
\item
  \textbf{Countdown}: Write a program that counts down from 10 to 1
  using \texttt{for}, \texttt{while} loops.
\item
  \textbf{Sum of Squares}: Write a function to calculate the sum of
  squares of all numbers from 1 to \texttt{n} using \texttt{for},
  \texttt{while} loops.
\item
  \textbf{Reverse String}: Write a function that reverses a given string
  using \texttt{while}, \texttt{for} loops.
\item
  \textbf{Matrix Transposition}: Write a function that transposes a
  given matrix (2D list) using nested \texttt{for}, \texttt{while}
  loops.
\item
  \textbf{Matrix Multiplication}: Write a function that performs matrix
  multiplication for two given matrices using nested \texttt{for} loops.
\item
  \textbf{Prime Factorization}: Write a function that returns the prime
  factors of a given number using a \texttt{for} loop and trial
  division.
\item
  \textbf{Palindrome Check}: Write a function that checks if a given
  string is a palindrome using a \texttt{for} loop.
\end{enumerate}

\clearpage

\section{Objects, Libraries, Data Science}
\subsubsection{Reflection from Last
Day:}\label{reflection-from-last-day}

\begin{itemize}

\item
  Answer questions about for iteration in Python
\item
  Discuss exercises from previous day
\end{itemize}

In Python, objects and classes are fundamental concepts used in
object-oriented programming (OOP). Object-oriented programming is a
programming paradigm that focuses on creating objects that encapsulate
data and behavior together. Here's an introduction to objects, classes,
and how they are used in Python:

\subsubsection{Objects}\label{objects}

An object is an instance of a class. It encapsulates data (attributes)
and behaviors (methods). In Python, almost everything is an object,
including integers, floats, strings, lists, and more complex structures
like functions and classes themselves.

\paragraph{Example of Objects:}\label{example-of-objects}

\begin{verbatim}
# Examples of objects
x = 5         # Integer object
name = "Alice"  # String object
my_list = [1, 2, 3]  # List object
\end{verbatim}

\subsubsection{Classes}\label{classes}

A class is a blueprint for creating objects (instances). It defines a
set of attributes and methods that characterize any object of the class.
Think of a class as a template or a blueprint that describes the
attributes and behaviors common to all objects of that class.

\paragraph{Defining a Class:}\label{defining-a-class}

In Python, a class is defined using the \texttt{class} keyword followed
by the class name and a colon \texttt{:}. Inside the class definition,
you define attributes and methods.

\paragraph{Example of a Class
Definition:}\label{example-of-a-class-definition}

\begin{verbatim}
# Class definition
class Person:
    # Constructor (initializing attributes)
    def __init__(self, name, age):
        self.name = name   # Attribute: name
        self.age = age     # Attribute: age
    
    # Method to greet
    def greet(self):
        return f"Hello, my name is {self.name} and I am {self.age} years old."

# Creating objects (instances) of the Person class
person1 = Person("Alice", 30)
person2 = Person("Bob", 25)

# Using object methods
print(person1.greet())  # Output: Hello, my name is Alice and I am 30 years old.
print(person2.greet())  # Output: Hello, my name is Bob and I am 25 years old.
\end{verbatim}

\paragraph{Anatomy of a Class:}\label{anatomy-of-a-class}

\begin{itemize}

\item
  \textbf{Attributes}: Variables that store data (state) related to the
  object (\texttt{self.name}, \texttt{self.age}).
\item
  \textbf{Methods}: Functions defined inside a class that can perform
  operations on the object (\texttt{greet()}).
\end{itemize}

\paragraph{\texorpdfstring{Constructor (\texttt{\_\_init\_\_}
method):}{Constructor (\_\_init\_\_ method):}}\label{constructor-__init__-method}

The \texttt{\_\_init\_\_} method is a special method in Python classes
that is called when an object is instantiated. It initializes the
object's attributes. The first parameter of \texttt{\_\_init\_\_} and
all instance methods is \texttt{self}, which refers to the instance of
the class itself.

\subsubsection{Key Concepts in OOP}\label{key-concepts-in-oop}

\begin{itemize}

\item
  \textbf{Encapsulation}: Bundling data (attributes) and methods
  (functions) that operate on the data into a single unit (object).
\item
  \textbf{Inheritance}: Creating new classes (derived classes or
  subclasses) from existing classes (base classes or superclasses) to
  reuse and extend functionality.
\item
  \textbf{Polymorphism}: The ability to use a single interface (method)
  to perform different tasks. In Python, polymorphism is achieved
  through method overriding and method overloading.
\end{itemize}

\subsubsection{Benefits of OOP}\label{benefits-of-oop}

\begin{itemize}

\item
  \textbf{Modularity}: Encapsulation allows for easier maintenance and
  modification of code.
\item
  \textbf{Reusability}: Classes and objects promote code reuse through
  inheritance.
\item
  \textbf{Flexibility and Scalability}: OOP makes it easier to manage
  and scale complex applications by organizing code into manageable
  pieces.
\end{itemize}

\subsubsection{Conclusion}\label{conclusion}

Understanding objects and classes in Python is crucial for leveraging
the power of object-oriented programming. By defining classes and
creating objects, you can organize your code more effectively, promote
code reuse, and build complex, scalable applications more efficiently.
OOP is a powerful paradigm that enhances code organization, readability,
and maintainability in Python and many other programming languages.

\subsection{Data Science Part 1}\label{data-science-part-1}

Data science is a growing field that combines mathematics, statistics,
and computer science to extract meaningful insights from data. By
analyzing and interpreting large amounts of information, data scientists
can uncover patterns, make predictions, and help solve complex problems.
In this introduction to data science, you will learn how to use Python
to analyze real-world data.

\paragraph{Learning Objectives}\label{learning-objectives}

By the end of today's session, you will have learned: - To ask good
questions about your data in preparation for analysis, - To install and
import a python library, - To read data into python from a url, - To
identify the level of observation of a dataset, - To explore a pandas
dataframe by applying attributes and methods on it.

\subsubsection{Python Libraries}\label{python-libraries}

A Python library is a collection of functions and methods that allows
you to perform specific tasks without having to write the code yourself
from scratch. Libraries are designed to be reusable, saving development
time and effort by providing pre-written solutions to common programming
challenges.

Python has a vast ecosystem of libraries covering various domains. Three
of the most popular libraries for data science projects are:

\begin{itemize}

\item
  \textbf{Pandas}: Data analysis library that provides high-performance
  data structures and tools for manipulating structured data.
\item
  \textbf{NumPy}: Numerical computing library for handling large arrays
  and matrices of numeric data.
\item
  \textbf{Matplotlib}: 2D plotting library for creating static,
  animated, and interactive visualizations.
\end{itemize}

You can learn about other popular python packages and their uses
\href{https://www.geeksforgeeks.org/python-libraries-to-know/}{here}.

\paragraph{1. Installing Libraries}\label{installing-libraries}

Before using a library, you need to install it. Python's package manager
\texttt{pip} is used for installing libraries from the Python Package
Index (PyPI).

\begin{verbatim}
pip install `library\_name`
\end{verbatim}

Replace \texttt{library\_name} with the name of the library you want to
install (e.g., \texttt{numpy}, \texttt{pandas}, \texttt{matplotlib}).

\paragraph{2. Importing Libraries}\label{importing-libraries}

Once installed, you import the library into your Python script or
interactive session using the \texttt{import} statement. You can also
use \texttt{import\ ...\ as\ ...} to create an alias for the library
name.

\subparagraph{Example: Importing NumPy and
Pandas}\label{example-importing-numpy-and-pandas}

\begin{verbatim}
import numpy as np
import pandas as pd
\end{verbatim}

\subsubsection{Japa!}\label{japa}

Every data science project begins with data, and understanding the
context behind the data is crucial for extracting meaningful insights.

Our dataset is extracted from the
\href{https://databank.worldbank.org/source/global-bilateral-migration}{World
Bank's} Global Bilateral Migration database. The dataset provides
detailed information on international migration patterns between
countries. It includes data on the number of migrants by origin and
destination countries, from 1960 to 2000, in ten-year intervals. This
dataset can help in analyzing migration trends, demographic changes, and
the impacts of migration on economies and societies globally. It is a
valuable resource for researchers, policymakers, and organizations
involved in migration studies and related fields.

Let's spend some time exploring the dataset.

\paragraph{Load the data}\label{load-the-data}

We begin by loading the data into Python using the \texttt{pandas}
library which offers many useful functions for reading data. Our dataset
is a CSV file stored in a GitHub repository; hence, we will use its URL
with the \texttt{read\_csv()} function to access it. (You can find more
examples on reading in csv files
\href{https://www.geeksforgeeks.org/reading-csv-files-in-python/}{here}.)

\begin{verbatim}
import pandas as pd

url = `https://raw.githubusercontent.com/naijacoderorg/lectures/main/
lectures2024/datascience/migrations.csv'
df = pd.read_csv(url)
\end{verbatim}

df is short form for DataFrame. It's a lazy way to name our initial
dataset. We should be more specific in our naming in the future.

\paragraph{Pandas DataFrame Attributes and
Methods}\label{pandas-dataframe-attributes-and-methods}

When we pull in data using pandas read.csv, the object we obtain
inherits a class of Pandas DataFrame.

\begin{verbatim}
df.__class__ # check for the class of df
\end{verbatim}

As a pandas dataframe, our \texttt{df} object has certain attributes and
methods. An \textbf{attribute} in programming is a property of an object
that stores data about that object. A \textbf{method} is a function that
can perform an action on an object of the specified class. Both
attributes and methods are accessed using dot notation, but methods
additionally have parentheses \texttt{()}---because they are
functions---even when no arguments are specified.

Let's explore some DataFrame attributes and methods for our dataset.

\paragraph{\texorpdfstring{\texttt{.head()} displays the first few rows
of the
DataFrame}{.head() displays the first few rows of the DataFrame}}\label{head-displays-the-first-few-rows-of-the-dataframe}

By default, only the first five rows are displayed. To get a different
number of rows, you can specify that in the argument, like
\texttt{df.head(10)}.

\begin{verbatim}
df.head()
\end{verbatim}

Our dataset is tabular. This means that it has rows and columns. The
rows represent a single observation, and the columns are features that
contain specific information (like country of origin) on each
observation.

Whenever we explore a dataset, one of the first questions we ask
ourselves is this: \textbf{What is the level of an observation?} Can you
describe what information a specific row gives you?

In this case, each row of the dataset represents a specific migration
flow over time. For example, the third row shows the number of
migrations from Afghanistan (the origin) to Algeria (the destination)
for the years 1960, 1970, 1980, 1990, and 2000. We always spend time
ensuring that we understand what each observation represents as the rest
of our analysis is hinged on this knowledge.

\paragraph{\texorpdfstring{\texttt{.shape} displays the number of rows
and
columns}{.shape displays the number of rows and columns}}\label{shape-displays-the-number-of-rows-and-columns}

We can also check for the dimensions (number of rows and columns) of our
dataset using the \texttt{.shape} attribute. Try to interpret the result
you get? Why do you think the dataset is this large? We'll explore this
later, but feel free to brainstorm now.

\begin{verbatim}
df.shape
\end{verbatim}

Try out the following attributes and methods to learn more about your
dataset. 
\begin{itemize}
\item df.columns
\item df.dtypes
\item df.index
\item df.size
\item df.values
\item df.info()
\item df.describe()
\item df.isnull().sum()
\end{itemize}

\subsubsection{Asking Good Data
Questions}\label{asking-good-data-questions}

In our next Data Science session, we will spend some time exploring the
data more deeply and finding exact answers to questions we have about
the data. But today, we need to learn how to generate effective data
questions that can lead to valuable insights and informed
decision-making.

\subsubsection{Principles for Formulating Good Data
Questions}\label{principles-for-formulating-good-data-questions}

\paragraph{1. Specificity}\label{specificity}

\begin{itemize}

\item
  Good questions are specific. Vague questions lead to unclear answers.
  Specific questions narrow down the focus and make it easier to find
  precise answers.
\item
  Example: Instead of asking ``How are our sales?'', ask ``How did our
  sales perform in the last quarter compared to the previous quarter?''
\end{itemize}

\paragraph{2. Measurability}\label{measurability}

\begin{itemize}

\item
  Questions should be measurable and quantifiable. If you can't measure
  it, you can't analyze it.
\item
  Example: ``What is the average customer satisfaction score for the
  past six months?'' rather than ``Are our customers happy?''
\end{itemize}

\paragraph{3. Relevance:}\label{relevance}

\begin{itemize}

\item
  Questions should be relevant to the business goals or research
  objectives. They should address a real need or problem.
\item
  Example: ``Which marketing channel has the highest return on
  investment in the past year?'' is more relevant than ``What are all
  the marketing channels we use?''
\end{itemize}

\paragraph{4. Actionability:}\label{actionability}

\begin{itemize}

\item
  A good data question leads to actionable insights. The answers should
  help make decisions or drive actions.
\item
  Example: ``What are the top three reasons customers are canceling
  their subscriptions?'' This can lead to actions to reduce churn.
\end{itemize}

In our work this week, we will ensure that our questions always satisfy
Principles 1 and 2. While our questions may not always meet Principles 3
and 4, these principles will become more useful as we begin practicing
asking data questions in real-world situations beyond this camp.

\subsubsection{Steps to Formulate Good Data
Questions:}\label{steps-to-formulate-good-data-questions}

\paragraph{1. Identify the Objective:}\label{identify-the-objective}

\begin{itemize}

\item
  Start by understanding the problem or goal. What do you want to
  achieve with this analysis?
\item
  Example: If the goal is to increase sales, your questions should focus
  on factors influencing sales.
\end{itemize}

\paragraph{2. Break Down the Problem:}\label{break-down-the-problem}

\begin{itemize}

\item
  Divide the main objective into smaller, more manageable questions.
\item
  Example: Instead of ``How can we increase sales?'', ask ``Which
  products have seen a decline in sales?'', ``Which customer segments
  are underperforming?'', and ``Which marketing campaigns were most
  effective?''
\end{itemize}

\paragraph{3. Consider the Data
Available:}\label{consider-the-data-available}

\begin{itemize}

\item
  Ensure that the questions can be answered with the data you have or
  can realistically obtain.
\item
  Example: If you have sales data by region, you can ask ``Which regions
  have the highest growth in sales?''
\end{itemize}

\paragraph{4. Iterate and Refine:}\label{iterate-and-refine}

\begin{itemize}

\item
  Continuously refine your questions based on initial findings and
  feedback.
\item
  Example: After analyzing the initial sales data, you might refine the
  question to focus on a specific product line or customer segment.
\end{itemize}

\subsubsection{Brainstorming Activity}\label{brainstorming-activity}

In small groups, let's practice coming up with good questions based on
our dataset. In your group, brainstorm a list of specific, measurable,
(relevant), and (actionable) questions that you would ask.

When the time is up, you will present your questions to the class.

\subsection{Exercises}\label{exercises}

\subsubsection{In the CLRS Textbook}\label{in-the-clrs-textbook}

\begin{enumerate}
\def\labelenumi{\arabic{enumi}.}

\item
  Read Chapter 1 (The Role of Algorithms in Computing) in CLRS.
\item
  Read Chapter 2 (Getting Started) in CLRS.
\item
  Read Chapter 3 (Growth of Functions) in CLRS.
\end{enumerate}

\subsubsection{Exercise 1: Using NumPy}\label{exercise-1-using-numpy}

\begin{enumerate}
\def\labelenumi{\arabic{enumi}.}

\item
  \textbf{Create a NumPy array}: Create a 1D array of integers from 1 to
  10.
\item
  \textbf{Calculate statistics}: Compute the mean, median, and standard
  deviation of the array.
\item
  \textbf{Reshape the array}: Reshape the array into a 2D matrix with 2
  rows and 5 columns.
\item
  \textbf{Perform element-wise operations}: Multiply each element of the
  matrix by 2.
\item
  \textbf{Indexing and slicing}: Extract the second row of the matrix.
\end{enumerate}

\subsubsection{Exercise 2: Using Pandas}\label{exercise-2-using-pandas}

\begin{enumerate}
\def\labelenumi{\arabic{enumi}.}

\item
  \textbf{Create a DataFrame}: Create a Pandas DataFrame from a
  dictionary with columns `Name', `Age', and `City'.
\item
  \textbf{Read from CSV}: Load a CSV file into a DataFrame using
  \texttt{pd.read\_csv()}.
\item
  \textbf{Data manipulation}: Add a new column `Salary' to the DataFrame
  with random salary values.
\item
  \textbf{Filtering and querying}: Filter rows where `Age' is greater
  than 30 and `City' is `New York'.
\item
  \textbf{Grouping and aggregation}: Group the data by `City' and
  calculate the average salary for each city.
\end{enumerate}

\subsubsection{Exercise 3: Using
Matplotlib}\label{exercise-3-using-matplotlib}

\begin{enumerate}
\def\labelenumi{\arabic{enumi}.}

\item
  \textbf{Line plot}: Plot a simple line graph of \texttt{y\ =\ sin(x)}
  for values of \texttt{x} from \(0\) to \(2\pi\).
\item
  \textbf{Scatter plot}: Generate random data points (x, y) and plot
  them as a scatter plot.
\item
  \textbf{Histogram}: Create a histogram of random data with 50 bins.
\item
  \textbf{Customizing plots}: Customize the plot by adding labels to
  axes, a title, and a legend (if applicable).
\item
  \textbf{Subplots}: Create a figure with multiple subplots, each
  displaying different types of plots (e.g., line plot, scatter plot).
\end{enumerate}

\subsubsection{Exercise 4: Combined
Exercises}\label{exercise-4-combined-exercises}

\begin{enumerate}
\def\labelenumi{\arabic{enumi}.}

\item
  \textbf{Data manipulation and visualization}: Load a dataset (e.g.,
  from Seaborn's built-in datasets or CSV), perform data manipulation
  using Pandas (e.g., filtering, grouping), and visualize the results
  using Matplotlib (e.g., bar plot, box plot).
\end{enumerate}

\subsubsection{Example Solutions}\label{example-solutions}

Here's a brief example of how you might approach Exercise 1 using NumPy:

\begin{verbatim}
import numpy as np

# Exercise 1: Using NumPy
# 1. Create a NumPy array
array = np.array([1, 2, 3, 4, 5, 6, 7, 8, 9, 10])

# 2. Calculate statistics
mean = np.mean(array)
median = np.median(array)
std_dev = np.std(array)

print(f"Mean: {mean}, Median: {median}, Standard Deviation: {std_dev}")

# 3. Reshape the array
matrix = array.reshape(2, 5)
print("Reshaped matrix:", matrix)

# 4. Perform element-wise operations
matrix *= 2
print("Matrix after multiplying by 2:\n", matrix)

# 5. Indexing and slicing
second_row = matrix[1, :]
print("Second row of matrix:\n", second_row)
\end{verbatim}

\section{Growth of Functions}
\subsubsection{Reflection from Last
Day:}\label{reflection-from-last-day}

\begin{itemize}

\item
  Answer questions about for objects, libraries in Python
\item
  Discuss exercises from previous day
\end{itemize}

\subsection{Data Science Part 2}\label{data-science-part-2}

\subsubsection{Reflection from Previous
Day}\label{reflection-from-previous-day}

\begin{itemize}

\item
  Recap of Day 1: Asking Good Questions and basic DataFrame inspections.
\end{itemize}

If you need to read in the migrations dataset again, here is the code
you need.

\begin{verbatim}
import pandas as pd

url = `https://raw.githubusercontent.com/naijacoderorg/lectures/main/
lectures2024/datascience/migrations.csv'

df = pd.read_csv(url)
df
\end{verbatim}

Last time, we learned how to ask good data questions. Today, we will
learn how to find answers to the following data questions. You can add
your own questions and follow the principles we learn today to answer
them.

\textbf{Q1. What were the top 10 migration routes in the world in each
of 1960, 1970, 1980, 1990, and 2000?}

\textbf{Q2. What were the top 10 migration routes ending in an African
country in each of 1960, 1970, 1980, 1990, and 2000?}

\textbf{Q3. What are the top 5 continents that Nigerians migrated to in
2000?}

\textbf{Q4. Ghana-Must-Go. Investigate this phenomenon by plotting a
line graph showing Nigeria-to-Ghana and Ghana-to-Nigeria migrations from
1960 to 2000. Then compare your results with the story in this
\href{https://atavist.mg.co.za/ghana-must-go-the-ugly-history-of-africas-most-famous-bag/}{article}.}

\paragraph{Learning Objectives}\label{learning-objectives}

By the end of today's session, you will have learned to: 
\begin{enumerate}
\item Manipulate dataframes to find answers to your data questions 
\item Select a column from a dataset 
\item Filter data based on some conditions
\item Sort a dataset based on the values of some columns
\item Group and aggregate data to calculate summary statistics for each group 
\item Generate plots to visualize your data
\end{enumerate}

\subsubsection{Concepts}\label{concepts}

To prepare us to answer our data questions, we will learn some important
concepts.

\emph{Note: \texttt{df} is our input dataset. We will use more
descriptive names for any new datasets we create.}

\paragraph{1. Selecting columns}\label{selecting-columns}

To select columns in a DataFrame, we pass a list of column names
enclosed within square brackets after the DataFrame name. Since a list
is also created using square brackets, this results in double square
brackets.

\begin{verbatim}
df[['dest_country']] # select a single column
df[['origin_country', 'dest_country']] # select multiple columns
\end{verbatim}

\paragraph{2. Sorting Data}\label{sorting-data}

Using the \texttt{sort\_values()} method, we can sort (arrange) the rows
of a DataFrame based on the values of one or more columns. This can be
useful if you want to present your results in ascending or descending
order.

Let's sort \texttt{df} \textbf{by} the values in the \texttt{1960}
column in \textbf{descending} order.

\begin{verbatim}
df.sort_values(by='1960', ascending=False)
\end{verbatim}

We can also sort by more than one column, which is useful when we want
to apply a secondary sorting criterion to break ties in the primary
sort. For example, if we want to sort our input dataset \texttt{df} by
the destination country first, but then sort by the origin country when
the destination values are the same, we can achieve this using
multi-column sorting.

\begin{verbatim}
# pass a list into the `by' argument to sort by multiple columns
df.sort_values(by=['dest_country', 'origin_country']) 
\end{verbatim}

We are now ready to answer our first question restated below:

\textbf{Q1. What were the top 10 migration routes in the world in each
of 1960, 1970, 1980, 1990, and 2000?}

First, the game plan: - Create a different table for each year - Select
only the necessary columns for each table: \texttt{origin\_country},
\texttt{dest\_country} and the year - Sort by values in the year column
in descending order - Display only the top 10 rows

Here is how we would answer this question for 1960.

\begin{verbatim}
top_immigrations_1960 = df[['origin_country', 'dest_country', '1960']]
.sort_values(by='1960', ascending=False).head(10)
top_immigrations_1960
\end{verbatim}

In the last code chunk, we sequentially applied column selections and
methods to our starting dataset. Sometimes, the code generated in this
process can get really long and messy. In such situations, we can break
it into several lines by wrapping the whole operation in parentheses.
This makes the code more readable while performing the same operations.

\begin{verbatim}
top_immigrations_1960b = (
    df[['origin_country', 'dest_country', '1960']]
    .sort_values(by='1960', ascending=False)
    .head(10)
)

top_immigrations_1960b
\end{verbatim}

\subparagraph{Practice Exercise 1}\label{practice-exercise-1}

Now, answer the question for the other years.

\paragraph{3. Filtering Data}\label{filtering-data}

Filtering in data analysis involves selecting rows from a DataFrame that
meet specific conditions. This allows you to focus on subsets of the
data that are relevant to your analysis. For example, you can filter
rows based on column values, such as selecting only the rows where a
particular column's value exceeds a certain threshold. This is done by
applying logical conditions directly to the DataFrame.

Let's say we want to filter the data to contain only rows that have
information on migrations FROM Ghana. What kind of logic should we
apply?

First of all, we need to find the column that contains information on
migration origins, which in this case is
\texttt{\textquotesingle{}origin\_country\textquotesingle{}}. We will
use this column to create a filtering criterion:
\texttt{\textquotesingle{}origin\_country\textquotesingle{}\ ==\ \textquotesingle{}Ghana\textquotesingle{}}.
Next, we use this criterion to apply a boolean condition on the dataset.

\begin{verbatim}
df[df['origin_country'] == 'Ghana']

# LET'S BREAK DOWN THIS CODE
### df['origin_country'] SELECTS A COLUMN
### df['origin_country'] == 'Ghana' CHECKS IF THE ROWS OF THAT COLUMN
### ARE EQUAL TO 'GHANA'. FOR EACH ROW, THIS RETURNS TRUE OR FALSE.
### df[df['origin_country'] == 'Ghana'] RETURNS ONLY ROWS OF df 
### THAT MEET THE CONDITION ABOVE
\end{verbatim}

\subparagraph{Practice Exercise 2}\label{practice-exercise-2}

\begin{itemize}

\item
  Create two datasets named \texttt{emigrations\_NGA} and
  \texttt{immigrations\_NGA} containing only rows that satisfy the
  description in the names.
\end{itemize}

Filtering on Multiple Conditions

We can also apply multiple conditions when filtering data. Let's create
a dataset named \texttt{nonzero\_emigrations\_NGA\_2000} containing only
rows of migrations from Nigeria with values greater than 0 in the year
2000. \emph{Note that for compound conditions, we put each condition in
parentheses to keep the logic clear.}

\begin{verbatim}
nonzero_emigrations_NGA_2000 = df[
    (df["origin_country"] == "Nigeria") &
    (df['2000'] > 0)
]

# note that all migrations in this dataset originate from Nigeria 
# and the 2000 column contains only nonzero values
nonzero_emigrations_NGA_2000 
\end{verbatim}

We are now ready to answer the second question restated below:

\textbf{Q2. What were the top 10 migration routes ending in an African
country in each of 1960, 1970, 1980, 1990, and 2000?}

Game plan: - Create a filtered dataset containing migrations to African
countries only - Select only the necessary columns for each table:
\texttt{origin\_country}, \texttt{dest\_country}, and the year - Sort by
values in the year column in descending order - Display only the top 10
rows

Here is an example for 1960. Do you notice anything interesting in the
results?

\begin{verbatim}
top_african_immigrations_1960 = (
    df[df["dest_continent"] == "Africa"]
    [['origin_country', 'dest_country', '1960']]
    .sort_values(by='1960', ascending=False)
    .head(10)
)

top_african_immigrations_1960
\end{verbatim}

\textbf{Practice Exercise 3}

Now, go ahead and create your own datasets for the other years.

\paragraph{4. Grouping and Aggregating
Data}\label{grouping-and-aggregating-data}

\textbf{Grouping Data}

Grouping data involves creating subsets of the DataFrame based on the
unique values in one or more columns. For instance, we might want to
group our migration dataset by the origin country.

\textbf{Aggregation Functions}

Aggregation functions are used to compute summary statistics for each
group. Common aggregation functions include:

\begin{itemize}

\item
  \textbf{Sum}: Adds up the values in each group.
\item
  \textbf{Mean}: Calculates the average of the values in each group.
\item
  \textbf{Count}: Counts the number of values in each group.
\item
  \textbf{Max}: Finds the maximum value in each group.
\item
  \textbf{Min}: Finds the minimum value in each group.
\end{itemize}

\textbf{Example}

If we want to find the sum of migrations for each origin country, we can
group \texttt{df} by \texttt{origin\_country} and then find the sum of
migrations for each year column.

In the code below, notice the sequential application of the
\texttt{.groupby()} and \texttt{.sum()} methods. We first apply
\texttt{.groupby()} and then apply the sum aggregation.

\begin{verbatim}
# Group data by origin and sum the migrations
emigrations_total = (df[['origin_country', '1960', '1970', '1980', '1990', '2000']]
                     .groupby('origin_country') #groupby
                     .sum() #aggregate
                     .reset_index()
                    )
emigrations_total
\end{verbatim}

We are ready to answer our third data question restated below:

\textbf{Q3. What are the top 5 continents that Nigerians migrated to in
2000?}

Game plan: - Filter for rows with origin country equal to Nigeria -
Group by origin country and destination continent - Select the `2000'
column before aggregating - Aggregate using a sum function - Reset
index. This step is necessary to add our grouping variables back into
the dataset as columns instead of as indices. - Sort values in
descending order

\begin{verbatim}
NGA_continent_emigrations = (df[df['origin_country'] == 'Nigeria']
                             .groupby(['origin_country','dest_continent'])
                             ['2000']
                             .sum()
                             .reset_index()
                             .sort_values(by='2000', ascending=False)
                            )

NGA_continent_emigrations
\end{verbatim}

\paragraph{5. Visualizing the data}\label{visualizing-the-data}

Visualization is key in data analysis because it turns complex numbers
into pictures that are easier to understand. By using charts and graphs,
we can quickly spot patterns and trends, making it easier to see what's
important. This helps us make better decisions because we can base them
on clear, visual information instead of just numbers.

We will use the matplotlib library for visualizing our data. Matplotlib
is a powerful and widely-used Python library for creating static,
animated, and interactive visualizations.

We start by importing matplotlib using an alias \emph{plt}.

\begin{verbatim}
import matplotlib.pyplot as plt
\end{verbatim}

To generate a plot, we write several lines of code, each line calling a
\textbf{pyplot} method (function) that serves a specific purpose. Here's
a brief explanation of the common methods used in
\texttt{matplotlib.pyplot\ (plt)} and why they are essential for
creating a visual:

\begin{itemize}
\item
  \texttt{plt.figure()}: This method creates a new figure or canvas for
  the plot. It's like setting up a blank sheet of paper where you'll
  draw your chart. You can also specify the size of the figure here
  using the \texttt{figsize} parameter.
\item
  Depending on the type of plot we want to create, we can use one of the
  following methods.

  \begin{itemize}
  \item
    \texttt{plt.plot()}: This is the core method for creating a line
    plot. You pass in the data you want to visualize, and plt.plot()
    draws the lines connecting your data points. It's the main action
    that puts the data on the canvas.
  \item
    \texttt{plt.scatter()}: Creates a scatter plot, which is used to
    show the relationship between two variables. It displays individual
    data points as dots, making it easier to see patterns or
    correlations.
  \item
    \texttt{plt.bar()}: Generates a bar chart, which is useful for
    comparing categorical data. Bars represent the frequency or amount
    of each category, allowing for easy comparison.
  \item
    \texttt{plt.hist()}: Creates a histogram, which is used to show the
    distribution of a dataset. It groups data into bins and displays the
    frequency of data points within each bin.
  \item
    \texttt{plt.boxplot()}: Produces a box plot, which visualizes the
    distribution of data based on quartiles. It highlights the median,
    quartiles, and outliers, providing insights into the data's spread
    and variability.
  \item
    \texttt{plt.pie()}: Makes a pie chart, which shows the proportion of
    each category as a slice of a pie. It's useful for illustrating how
    different parts make up a whole.
  \item
    \texttt{plt.barh()}: Creates a horizontal bar chart, similar to
    plt.bar() but with bars oriented horizontally. This can be useful
    for displaying longer category names.
  \end{itemize}
\item
  \texttt{plt.title()}: This method adds a title to your plot. It's
  important because it gives context to the visual, letting the viewer
  know what the data represents.
\item
  \texttt{plt.xlabel()} and \texttt{plt.ylabel()}: These methods label
  the x-axis and y-axis, respectively. Labels are crucial because they
  tell the viewer what each axis represents, making the chart easier to
  understand.
\item
  \texttt{plt.legend()}: When you have multiple lines or data series in
  one plot, the legend helps identify which line corresponds to which
  data. It adds a small box that labels each line, making the plot
  clearer.
\item
  \texttt{plt.show()}: This method displays the plot. Without it, the
  plot might not appear, especially in some programming environments.
  It's like hitting the ``display'' button to see the final chart.
\end{itemize}

Each of these methods contributes to creating a clear, informative, and
visually appealing plot that effectively communicates the data to the
viewer.

\subparagraph{Example}\label{example}

Using a bar graph, visualize the number of migrations from Nigeria in
2000 by destination continent. \emph{Hint:} Use the dataset you created
for Q3.

\begin{verbatim}
# Create the bar chart
plt.figure(figsize=(10, 6))  # Optional: Set the figure size
plt.bar(NGA_continent_emigrations['dest_continent'], 
NGA_continent_emigrations['2000'], 
color='green')
# first arg is for x-axis, second arg is for y-axis, 
# we can optionally specify a color for the bars

# Add labels and title
plt.xlabel('Destination Continent')
plt.ylabel('Number of Emigrations')
plt.title('Migrations from Nigeria in 2000 by Destination Continent')

# Display the plot
plt.show()
\end{verbatim}

You can now attempt Q4 restated below using the concepts we have learned
today.

\textbf{Q4. Ghana-Must-Go. Investigate this phenomenon by plotting a
line graph showing Nigeria-to-Ghana and Ghana-to-Nigeria migrations from
1960 to 2000. Then compare your results with the story in this
\href{https://atavist.mg.co.za/ghana-must-go-the-ugly-history-of-africas-most-famous-bag/}{article}.}

\subsection{Growth of Functions}\label{growth-of-functions}

Understanding the growth of functions and asymptotic notation is crucial
in analyzing the efficiency of algorithms and data structures. These
concepts help in quantifying how the runtime or space requirements of
algorithms change with input size. Let's get into each of these
concepts.

In computer science, the growth of a function refers to how its runtime
(or space) increases as the size of the input grows. This growth is
typically measured in terms of the input size \(n\).

\paragraph{Example of Growth
Functions:}\label{example-of-growth-functions}

Consider two functions:

\begin{enumerate}
\item $f(n) = 2n + 1$
\item $g(n) = n^2 + 3n$
\end{enumerate}

As \(n\) increases:

\begin{enumerate}
\item $f(n)$ grows linearly $O(n)$ because the dominant term is $n$.
\item $g(n)$ grows quadratically $O(n^2)$ because the dominant term is $n^2$.
\end{enumerate}

\subsubsection{Asymptotic Notation}

Asymptotic notation provides a formal way to describe the limiting
behavior (as \(n\) becomes very large) of a function:

\begin{enumerate}

\item \textbf{Big O Notation (O}:
\begin{enumerate}
\item \textbf{Definition}: Represents the upper bound of the function's growth rate. It describes the worst-case scenario.
\item \textbf{Example}: Intuition is that $O(n^2)$ means the function grows no faster than $n^2$.
\end{enumerate}

\item \textbf{Omega Notation ($\Omega$)}:
\begin{enumerate}
\item \textbf{Definition}: Represents the lower bound of the function's growth rate. It describes the best-case scenario.
\item \textbf{Example}: Intuition is that $\Omega(n)$ means the function grows at least as fast as $n$.
\end{enumerate}

\item \textbf{Theta Notation ($\Theta$)}:
\begin{enumerate}
\item \textbf{Definition}: Represents both the upper and lower bounds, providing a tight bound on the function's growth rate.
\item \textbf{Example}: Intuition is that $\Theta(n)$ means the function grows exactly like $n$.
\end{enumerate}

\end{enumerate}

In mathematical terms, Big O notation \(O(f(n))\) is used to denote an
upper bound on the asymptotic growth rate of a function \(f(n)\) as
\(n\) becomes large. Here's the mathematical definition of Big O
notation:

\subsubsection{Definition of Big O}\label{definition-of-big-o}

Given two functions \(f(n)\) and \(g(n)\):

\[f(n) = O(g(n))\]

This statement means that there exist constants \(c > 0\) and
\(n_0 \geq 0\) such that for all \(n \geq n_0\), the function \(f(n)\)
is bounded above by \(c \cdot g(n)\):

\[|f(n)| \leq c \cdot |g(n)| \quad \text{for all } n \geq n_0\]

\paragraph{Key Points}\label{key-points}

\begin{itemize}
\item\textbf{Upper Bound}: Big O notation provides an upper bound on the growth rate of $f(n)$.
\item\textbf{Constant Factors}: It ignores constant factors and lower-order terms, focusing on the dominant term that contributes most significantly to the growth rate as $n$ increases.
\item\textbf{Asymptotic Behavior}: Big O notation describes the long-term behavior of functions as $n$ approaches infinity.
\end{itemize}

\subsubsection{Example Interpretation}\label{example-interpretation}

If \(f(n) = 2n + 1\), we say \(f(n) = O(n)\) because for sufficiently
large \(n\), \(2n + 1\) is bounded above by \(cn\) for some constant
\(c\). Specifically, for \(c = 3\) and \(n_0 = 1\), \(2n + 1 \leq 3n\)
holds for all \(n \geq 1\).

\subsubsection{Practical Use}\label{practical-use}

In algorithm analysis, Big O notation is widely used to characterize the
time complexity (and sometimes space complexity) of algorithms. It helps
in comparing the efficiency of algorithms and predicting how they will
perform with larger inputs.

In mathematical terms, Big Omega notation \(\Omega(f(n))\) is used to
denote a lower bound on the asymptotic growth rate of a function
\(f(n)\) as \(n\) becomes large. Here's the mathematical definition of
Big Omega notation:

\subsubsection{Definition of Big Omega}\label{definition-of-big-omega}

Given two functions \(f(n)\) and \(g(n)\):

\[f(n) = \Omega(g(n))\]

This statement means that there exist constants \(c > 0\) and
\(n_0 \geq 0\) such that for all \(n \geq n_0\), the function \(f(n)\)
is bounded below by \(c \cdot g(n)\):

\[|f(n)| \geq c \cdot |g(n)| \quad \text{for all } n \geq n_0\]

\paragraph{Key Points}\label{key-points-1}

\begin{itemize}
\item\textbf{Lower Bound}: Big Omega notation provides a lower bound on the growth rate of $f(n)$.
\item\textbf{Constant Factors}: Similar to Big O notation, it ignores constant factors and lower-order terms, focusing on the dominant term that contributes most significantly to the growth rate as $n$ increases.
\item\textbf{Asymptotic Behavior}: Big Omega notation describes the long-term behavior of functions as $n$ approaches infinity.
\end{itemize}

\subsubsection{Example Interpretation}\label{example-interpretation-1}

If \(f(n) = n^2 + 3n\), we say \(f(n) = \Omega(n^2)\) because for
sufficiently large \(n\), \(n^2 + 3n\) is bounded below by \(cn^2\) for
some constant \(c\). Specifically, for \(c = 1\) and \(n_0 = 1\),
\(n^2 + 3n \geq n^2\) holds for all \(n \geq 1\).

\subsubsection{Practical Use}\label{practical-use-1}

Big Omega notation is used similarly to Big O notation in algorithm
analysis. It helps in characterizing the lower bounds of algorithms'
time complexity (and sometimes space complexity), providing insight into
their performance guarantees for large inputs.

\subsubsection{Comparison with Big O
Notation}\label{comparison-with-big-o-notation}

\textbf{Big O vs. Big Omega}: While Big O provides an upper bound on the
growth rate of a function \(f(n)\), Big Omega provides a lower bound.
Together, they can give a more complete picture of an algorithm's
complexity by indicating both its worst-case and best-case scenarios.

In mathematical terms, Big Theta notation \(\Theta(f(n))\) is used to
denote a tight bound on the asymptotic growth rate of a function
\(f(n)\) as \(n\) becomes large. Here's the mathematical definition of
Big Theta notation:

\subsubsection{Definition of Big Theta}\label{definition-of-big-theta}

Given two functions \(f(n)\) and \(g(n)\):

\[f(n) = \Theta(g(n))\]

This statement means that there exist constants \(c_1 > 0\),
\(c_2 > 0\), and \(n_0 \geq 0\) such that for all \(n \geq n_0\), the
function \(f(n)\) is bounded both above and below by \(c_1 \cdot g(n)\)
and \(c_2 \cdot g(n)\), respectively:

\[c_1 \cdot |g(n)| \leq |f(n)| \leq c_2 \cdot |g(n)| \quad \text{for all } n \geq n_0\]

\paragraph{Key Points}\label{key-points-2}

\begin{itemize}
\item\textbf{Tight Bound}: Big Theta notation provides a tight or asymptotically tight bound on the growth rate of $f(n)$. It means $f(n)$ grows at the same rate as $g(n)$ within constant factors, asymptotically speaking.
  
\item\textbf{Constant Factors}: Similar to Big O and Big Omega notations, Big Theta ignores constant factors and lower-order terms, focusing on the dominant term that contributes most significantly to the growth rate as $n$ increases.
  
\item\textbf{Asymptotic Behavior}: Big Theta notation describes the long-term behavior of functions as $n$ approaches infinity, indicating that $f(n)$ and $g(n)$ grow at the same rate.
\end{itemize}

\subsubsection{Example Interpretation}\label{example-interpretation-2}

If \(f(n) = n^2 + 3n\), and \(g(n) = n^2\), we say
\(f(n) = \Theta(n^2)\) because there exist constants \(c_1 = 1\),
\(c_2 = 1\), and \(n_0 = 1\) such that \(n^2 \leq n^2 + 3n \leq n^2\)
holds for all \(n \geq 1\).

\subsubsection{Practical Use}\label{practical-use-2}

Big Theta notation is particularly useful in algorithm analysis as it
provides a precise description of an algorithm's asymptotic behavior. It
indicates both the upper and lower bounds of the growth rate of
algorithms' time complexity (and sometimes space complexity), giving a
clear understanding of their efficiency across different input sizes.

\subsubsection{Comparison with Big O and Big Omega
Notations}\label{comparison-with-big-o-and-big-omega-notations}

\begin{itemize}
\item Big O provides an upper bound on the growth rate of $f(n)$.
\item Big Omega provides a lower bound on the growth rate of $f(n)$.
\item Big Theta provides both upper and lower bounds, indicating that $f(n)$ grows asymptotically at the same rate as $g(n)$.
\end{itemize}

In mathematical analysis, small o \(o(f(n))\) and small omega
\(\omega(f(n))\) notations are used to describe the asymptotic behavior
of functions in terms of limits. Here's how they are defined using
limits:

\subsubsection{Small o Notation}\label{small-o-notation}

Small o notation \(o(f(n))\) denotes that a function \(g(n)\) grows
asymptotically slower than \(f(n)\) as \(n\) approaches infinity.

\paragraph{Definition of Small o}\label{definition-of-small-o}

\[g(n) = o(f(n)) \quad \text{if and only if} \quad \lim_{n \to \infty} \frac{g(n)}{f(n)} = 0\]

In simpler terms, \(g(n)\) is in \(o(f(n))\) if for every positive
constant \(c > 0\), there exists a constant \(n_0 \geq 0\) such that for
all \(n \geq n_0\), \(|g(n)| < c \cdot |f(n)|\).

\subsubsection{Small Omega Notation}\label{small-omega-notation}

Small omega notation \(\omega(f(n))\) denotes that a function \(g(n)\)
grows asymptotically faster than \(f(n)\) as \(n\) approaches infinity.

\paragraph{Definition of Small Omega}\label{definition-of-small-omega}

\[g(n) = \omega(f(n)) \quad \text{if and only if} \quad \lim_{n \to \infty} \frac{g(n)}{f(n)} = \infty\]

In other words, \(g(n)\) is in \(\omega(f(n))\) if for every positive
constant \(c > 0\), there exists a constant \(n_0 \geq 0\) such that for
all \(n \geq n_0\), \(|g(n)| > c \cdot |f(n)|\).

\paragraph{Key Points}\label{key-points-3}

\begin{itemize}
\item\textbf{Asymptotic Behavior}: Small o notation $o(f(n))$ indicates that $g(n)$ grows slower than $f(n)$ as $n$ becomes large, while small omega $\omega(f(n))$ indicates that $g(n)$ grows faster than $f(n)$.
  
\item\textbf{Limits}: Both notations are defined using limits to formalize the concept of growth rate in asymptotic analysis.
  
\item\textbf{Precision}: These notations provide a more precise characterization of asymptotic relationships compared to Big O and Big Omega, especially in contexts where exact growth rates are important.
\end{itemize}

\subsection{Exercises for Asymptotic Notation
Section}\label{exercises-for-asymptotic-notation-section}

\subsubsection{Exercise 1: Identifying
Notations}\label{exercise-1-identifying-notations}

For each pair of functions \(f(n)\) and \(g(n)\), determine whether
\(f(n)\) is in Big O, Small o, Big Theta, Small omega, or Big Omega
relative to \(g(n)\).

\begin{enumerate}
\item $f(n) = 2n + 1$, $g(n) = n$
\item $f(n) = n^2$, $g(n) = 3n^2$
\item $f(n) = \log n$, $g(n) = \sqrt{n}$
\item $f(n) = n^{0.5}$, $g(n) = n^{0.4}$
\item $f(n) = n \log n$, $g(n) = n$
\end{enumerate}

\subsubsection{Exercise 2: Verifying
Limits}\label{exercise-2-verifying-limits}

Determine whether the following statements are true (T) or false (F):

\begin{enumerate}
\item $n^2 = O(n^3)$
\item $n^2 = o(n^3)$
\item $n^2 = \Theta(n^2)$
\item $n^2 = \omega(n)$
\item $n^2 = \Omega(n^2)$
\end{enumerate}

\subsubsection{Exercise 3: Matching
Functions}\label{exercise-3-matching-functions}

Match each function \(f(n)\) with the appropriate notation (Big O, Small
o, Big Theta, Small omega, Big Omega) describing its asymptotic
behavior:

\begin{enumerate}
\item $f(n) = n^2$
\item $f(n) = \log n$
\item $f(n) = 3n + 5$
\item $f(n) = n^{0.5}$
\item $f(n) = 2^n$
\end{enumerate}

\clearpage

\section{Searching Algorithms}
\subsubsection{Reflection from Last
Day:}\label{reflection-from-last-day}

\begin{itemize}

\item
  Answer questions about for asymptotic notation and growth of functions
\item
  Discuss exercises from previous day
\end{itemize}

Searching algorithms are fundamental methods used to locate elements
within a data structure, such as arrays, lists, trees, graphs, and more.
In Python, these algorithms are crucial for efficiently finding specific
items based on certain criteria. Here's an introduction to some commonly
used searching algorithms:

\subsubsection{Linear Search}\label{linear-search}

\begin{itemize}
\item\textbf{Linear search} is the simplest form of searching algorithm where each element in the list is checked sequentially until the target element is found or the end of the list is reached.
\item\textbf{Time Complexity}: $O(n)$ - Linear time complexity, where $n$ is the number of elements in the list.
\end{itemize}

\paragraph{Example of Linear Search in
Python:}\label{example-of-linear-search-in-python}

\begin{verbatim}
def linear_search(arr, target):
    for i in range(len(arr)):
        if arr[i] == target:
            return i  # Return index if found
    return -1  # Return -1 if not found

# Example usage
my_list = [10, 30, 20, 5, 15]
target_value = 20
result = linear_search(my_list, target_value)
if result != -1:
    print(f"Element found at index {result}")
else:
    print("Element not found")
\end{verbatim}

\subsubsection{Binary Search}\label{binary-search}

\textbf{Binary search} is a more efficient searching algorithm that
requires the list to be sorted. It works by repeatedly dividing the
search interval in half. If the target value matches the middle element,
its position is returned. Otherwise, the search continues in either the
left or right half, depending on whether the target value is less than
or greater than the middle element.

\textbf{Time Complexity}: \(O(\log n)\) - Logarithmic time complexity on
a sorted list (where n is the number of elements in the sorted list).

\paragraph{Example of Binary Search in
Python:}\label{example-of-binary-search-in-python}

\begin{verbatim}
def binary_search(arr, target):
    low = 0
    high = len(arr) - 1
    while low <= high:
        mid = (low + high) // 2
        if arr[mid] == target:
            return mid  # Return index if found
        elif arr[mid] < target:
            low = mid + 1  # Search in the right half
        else:
            high = mid - 1  # Search in the left half
    return -1  # Return -1 if not found

# Example usage
sorted_list = [5, 10, 15, 20, 30]
target_value = 20
result = binary_search(sorted_list, target_value)
if result != -1:
    print(f"Element found at index {result}")
else:
    print("Element not found")
\end{verbatim}

\subsubsection{Key Considerations}\label{key-considerations}

\begin{itemize}

\item
  \textbf{Data Structure}: The choice of searching algorithm often
  depends on the data structure being used (e.g., lists, trees, graphs).
\item
  \textbf{Performance}: Binary search is significantly faster than
  linear search for large datasets, especially when the data is sorted.
\item
  \textbf{Edge Cases}: Consider edge cases such as empty lists or arrays
  with duplicate values when implementing and testing searching
  algorithms.
\end{itemize}

\subsection{Exercises}\label{exercises}

\subsubsection{Linear Search Exercises}\label{linear-search-exercises}

\begin{enumerate}
\def\labelenumi{\arabic{enumi}.}

\item
  \textbf{Exercise 1: Finding an Element}

  \begin{itemize}
  \item
    \textbf{Problem}: Implement a function
    \texttt{linear\_search(arr,\ target)} that returns the index of
    \texttt{target} in the list \texttt{arr} using linear search. If
    \texttt{target} is not present, return \texttt{-1}.
  \item
    \textbf{Example}:

\begin{verbatim}
arr = [3, 1, 4, 1, 5, 9, 2, 6, 5, 3]
target = 5
# Output: 4 (index of the first occurrence of 5)
\end{verbatim}

    Use a while loop!
  \end{itemize}
\item
  \textbf{Exercise 2: Counting Occurrences}

  \begin{itemize}
  \item
    \textbf{Problem}: Modify the previous function to return the number
    of times \texttt{target} appears in \texttt{arr}.
  \item
    \textbf{Example}:

\begin{verbatim}
arr = [3, 1, 4, 1, 5, 9, 2, 6, 5, 3]
target = 5
# Output: 2 (number of times 5 appears in arr)
\end{verbatim}
  \end{itemize}
\item
  \textbf{Exercise 3: Sum of Elements}

  \begin{itemize}
  \item
    \textbf{Problem}: Write a function \texttt{sum\_linear\_search(arr)}
    that computes the sum of all elements in the list \texttt{arr} using
    linear search.
  \item
    \textbf{Example}:

\begin{verbatim}
arr = [3, 1, 4, 1, 5, 9, 2, 6, 5, 3]
# Output: 39 (sum of all elements in arr)
\end{verbatim}
  \end{itemize}
\end{enumerate}

\subsubsection{Binary Search Exercises}\label{binary-search-exercises}

\begin{enumerate}
\def\labelenumi{\arabic{enumi}.}

\item
  \textbf{Exercise 4: Basic Binary Search}

  \begin{itemize}
  \item
    \textbf{Problem}: Implement a function
    \texttt{binary\_search(arr,\ target)} that performs binary search on
    a sorted list \texttt{arr} to find the index of \texttt{target}. If
    \texttt{target} is not present, return \texttt{-1}.
  \item
    \textbf{Example}:

\begin{verbatim}
arr = [1, 2, 3, 4, 5, 6, 7, 8, 9, 10]
target = 6
# Output: 5 (index of 6 in arr)
\end{verbatim}

    Use a for loop or recursion!
  \end{itemize}
\item
  \textbf{Exercise 5: Finding Smallest Element}

  \begin{itemize}
  \item
    \textbf{Problem}: Modify the binary search function to find the
    smallest element in a rotated sorted array \texttt{rotated\_arr}.
  \item
    \textbf{Example}:

\begin{verbatim}
rotated_arr = [4, 5, 6, 7, 0, 1, 2]
# Output: 0 (smallest element in rotated_arr)
\end{verbatim}
  \end{itemize}
\item
  \textbf{Exercise 6: Counting Elements}

  \begin{itemize}
  \item
    \textbf{Problem}: Write a function
    \texttt{count\_binary\_search(arr,\ target)} that counts the number
    of occurrences of \texttt{target} in a sorted list \texttt{arr}
    using binary search.
  \item
    \textbf{Example}:

\begin{verbatim}
arr = [1, 2, 2, 2, 3, 4, 5, 5, 6]
target = 2
# Output: 3 (number of times 2 appears in arr)
\end{verbatim}
  \end{itemize}
\end{enumerate}

\clearpage

\section{Sorting Algorithms: Bubble, Selection, Insertion}
\subsubsection{Reflection from Last
Day:}\label{reflection-from-last-day}

\begin{itemize}

\item
  Answer questions about searching algorithms
\item
  Discuss exercises from previous day
\end{itemize}

Sorting algorithms are fundamental to computer science and are used to
rearrange elements in a list or array into a specified order, typically
numerical or lexicographical. There are numerous sorting algorithms,
each with its own characteristics in terms of complexity, stability, and
suitability for different data sizes and types. Here's an introduction
to some common sorting algorithms:

\begin{enumerate}
\item \textbf{Bubble Sort}
\begin{itemize}
\item\textbf{Description}: Bubble Sort repeatedly steps through the list, compares adjacent elements, and swaps them if they are in the wrong order. The pass through the list is repeated until the list is sorted.
\item\textbf{Complexity}: $O(n^2)$ in the worst-case scenario.
\item\textbf{Key Features}: Simple implementation, but inefficient for large data sets.
\end{itemize}

\item \textbf{Selection Sort}
\begin{itemize}
\item\textbf{Description}: Selection Sort divides the list into sorted and unsorted parts. It repeatedly selects the smallest (or largest) element from the unsorted part and swaps it with the first unsorted element.
\item\textbf{Complexity}: $O(n^2)$ in all cases.
\item\textbf{Key Features}: Simple implementation, but also inefficient for large data sets.
\end{itemize}

\item \textbf{Insertion Sort}
\begin{itemize}
\item\textbf{Description}: Insertion Sort builds the final sorted array one item at a time. It takes each element from the list and inserts it into its correct position relative to the sorted portion of the array.
\item\textbf{Complexity}: $O(n^2)$ in the worst-case scenario, but $O(n)$ when the list is nearly sorted.
\item\textbf{Key Features}: Efficient for small data sets or when the input array is almost sorted.
\end{itemize}

\item \textbf{Merge Sort}
\begin{itemize}
\item\textbf{Description}: Merge Sort is a divide-and-conquer algorithm. It divides the input array into two halves, recursively sorts each half, and then merges the sorted halves to produce the final sorted array.
\item\textbf{Complexity}: $O(n \log n)$ in all cases.
\item\textbf{Key Features}: Stable sort (keeps the order of equal elements), efficient for large data sets, but requires additional memory proportional to the size of the input.
\end{itemize}

\item \textbf{Quick Sort}
\begin{itemize}
\item\textbf{Description}: Quick Sort is another divide-and-conquer algorithm. It picks an element as a pivot and partitions the array around the pivot, such that elements less than the pivot are on the left and elements greater than the pivot are on the right. It then recursively sorts the sub-arrays.
\item\textbf{Complexity}: $O(n \log n)$ in the average case, $O(n^2)$ in the worst-case scenario (rare).
\item\textbf{Key Features}: Efficient for large data sets, in-place (requires only a small additional memory), and often faster than Merge Sort in practice.
\end{itemize}

\end{enumerate}

\subsubsection{Choosing the Right
Algorithm}\label{choosing-the-right-algorithm}

\begin{itemize}

\item
  \textbf{Efficiency}: Consider the size of the data set and the
  expected range of data values.
\item
  \textbf{Stability}: Whether the algorithm maintains the relative order
  of equal elements.
\item
  \textbf{Memory Usage}: Some algorithms require additional memory for
  operations.
\item
  \textbf{Application}: Different algorithms may be more suitable
  depending on the specific requirements and characteristics of the
  data.
\end{itemize}

Each sorting algorithm has its trade-offs in terms of time complexity,
space complexity, and practical performance characteristics. The choice
of sorting algorithm depends on the specific requirements of the
application, such as input size, data characteristics, stability, and
performance constraints.

\subsection{Insertion Sort}\label{insertion-sort}

Insertion Sort is a simple sorting algorithm that builds the final
sorted array (or list) one element at a time. It is efficient for small
data sets or when the input array is almost sorted. Here's an
introduction to Insertion Sort:

\subsubsection{Description}\label{description}

Insertion Sort works similarly to how you might sort playing cards in
your hands. You start with an empty left hand and the cards face down on
the table. You then remove one card at a time from the table and insert
it into the correct position in your left hand, maintaining the cards in
your left hand sorted.

\subsubsection{Steps}\label{steps}

\begin{enumerate}
\def\labelenumi{\arabic{enumi}.}
\item
  \textbf{Initialization}: Start with the second element (index 1) and
  consider it as part of the sorted portion.
\item
  \textbf{Comparison and Insertion}: For each element, compare it with
  the elements in the sorted portion (left side). Shift all the elements
  greater than the current element to the right. Insert the current
  element into its correct position.
\item
  \textbf{Repeat}: Repeat the process until the entire array is sorted.
\end{enumerate}

\subsubsection{Example}\label{example}

Let's sort the array \texttt{{[}5,\ 2,\ 4,\ 6,\ 1,\ 3{]}} using
Insertion Sort:

\begin{itemize}

\item
  \textbf{Initial Array}: \texttt{{[}5,\ 2,\ 4,\ 6,\ 1,\ 3{]}}
\end{itemize}

\begin{enumerate}
\def\labelenumi{\arabic{enumi}.}
\item
  Start with the second element (\texttt{2}):

  \begin{itemize}
  
  \item
    Compare \texttt{2} with \texttt{5} (first element). Since
    \texttt{2\ \textless{}\ 5}, swap them.
  \item
    Array becomes \texttt{{[}2,\ 5,\ 4,\ 6,\ 1,\ 3{]}}.
  \end{itemize}
\item
  Consider the third element (\texttt{4}):

  \begin{itemize}
  
  \item
    Compare \texttt{4} with \texttt{5} (previous element). Since
    \texttt{4\ \textless{}\ 5}, swap them.
  \item
    Array becomes \texttt{{[}2,\ 4,\ 5,\ 6,\ 1,\ 3{]}}.
  \item
    Compare \texttt{4} with \texttt{2} (previous element). No swap
    needed.
  \end{itemize}
\item
  Continue with each subsequent element, shifting larger elements as
  necessary and inserting the current element into its correct position.
\item
  After sorting, the array becomes \texttt{{[}1,\ 2,\ 3,\ 4,\ 5,\ 6{]}}.
\end{enumerate}

\subsubsection{Complexity}\label{complexity}

\begin{itemize}
\item\textbf{Time Complexity}: $O(n^2)$ in the worst-case scenario (when the array is reverse sorted), and $O(n)$ in the best-case scenario (when the array is already sorted).
\item\textbf{Space Complexity}: $O(1)$ additional space, as it sorts in-place.
\end{itemize}

\subsubsection{Implementation (Python)}\label{implementation-python}

Here's a simple implementation of Insertion Sort in Python:

\begin{verbatim}
def insertion_sort(arr):
    for i in range(1, len(arr)):
        current_value = arr[i]
        j = i - 1
        while j >= 0 and arr[j] > current_value:
            arr[j + 1] = arr[j]
            j -= 1
        arr[j + 1] = current_value

# Example usage:
arr = [5, 2, 4, 6, 1, 3]
insertion_sort(arr)
print("Sorted array:", arr)
\end{verbatim}

In this implementation: - We iterate over each element in the array
starting from index 1. - For each element, we compare it with the
elements to its left in the sorted portion of the array and insert it
into its correct position by shifting larger elements to the right.

Insertion Sort is intuitive and straightforward to implement, making it
suitable for small data sets or scenarios where the input is mostly
sorted. However, for large data sets, more efficient algorithms like
Merge Sort or Quick Sort are generally preferred due to their better
average-case performance.

\subsection{Selection Sort}\label{selection-sort}

Selection Sort is a simple and intuitive sorting algorithm that divides
the input list into a sorted and an unsorted region. It repeatedly
selects the smallest (or largest, depending on the sorting order)
element from the unsorted region and swaps it with the first element of
the unsorted region. Here's an introduction to Selection Sort:

\subsubsection{Description}\label{description-1}

Selection Sort works by finding the minimum element from the unsorted
portion of the list and swapping it with the first unsorted element. It
continues this process, gradually reducing the unsorted region until the
entire list is sorted.

\subsubsection{Steps}\label{steps-1}

\begin{enumerate}
\def\labelenumi{\arabic{enumi}.}
\item
  \textbf{Initialization}: Start with the entire list considered as
  unsorted.
\item
  \textbf{Finding the Minimum}: Iterate through the unsorted region to
  find the minimum element.
\item
  \textbf{Swapping}: Swap the minimum element with the first element of
  the unsorted region.
\item
  \textbf{Repeat}: Continue the process for the remaining unsorted
  region until the list is sorted.
\end{enumerate}

\subsubsection{Example}\label{example-1}

Let's sort the array \texttt{{[}64,\ 25,\ 12,\ 22,\ 11{]}} using
Selection Sort:

\begin{itemize}

\item
  \textbf{Initial Array}: \texttt{{[}64,\ 25,\ 12,\ 22,\ 11{]}}
\end{itemize}

\begin{enumerate}
\def\labelenumi{\arabic{enumi}.}

\item
  \textbf{First Pass}: Find the smallest element and swap it with the
  first element:

  \begin{itemize}
  
  \item
    \texttt{{[}11,\ 25,\ 12,\ 22,\ 64{]}} (swap \texttt{11} with
    \texttt{64})
  \end{itemize}
\item
  \textbf{Second Pass}: Consider the remaining unsorted region
  (\texttt{{[}25,\ 12,\ 22,\ 64{]}}):

  \begin{itemize}
  
  \item
    \texttt{{[}11,\ 12,\ 25,\ 22,\ 64{]}} (swap \texttt{12} with
    \texttt{25})
  \end{itemize}
\item
  \textbf{Third Pass}: Continue with the remaining unsorted region
  (\texttt{{[}25,\ 22,\ 64{]}}):

  \begin{itemize}
  
  \item
    \texttt{{[}11,\ 12,\ 22,\ 25,\ 64{]}} (swap \texttt{22} with
    \texttt{25})
  \end{itemize}
\item
  \textbf{Fourth Pass}: Continue with the remaining unsorted region
  (\texttt{{[}64{]}}):

  \begin{itemize}
  
  \item
    \texttt{{[}11,\ 12,\ 22,\ 25,\ 64{]}} (no swap needed)
  \end{itemize}
\item
  \textbf{Final Sorted Array}: \texttt{{[}11,\ 12,\ 22,\ 25,\ 64{]}}
\end{enumerate}

\subsubsection{Complexity}\label{complexity-1}

\begin{itemize}
\item\textbf{Time Complexity}: $O(n^2)$ in all cases (worst-case, average-case, and best-case scenarios).
\item\textbf{Space Complexity}: $O(1)$ additional space, as it sorts in-place.
\end{itemize}

\subsubsection{Implementation (Python)}\label{implementation-python-1}

Here's a simple implementation of Selection Sort in Python:

\begin{verbatim}
def selection_sort(arr):
    n = len(arr)
    for i in range(n):
        # Find the minimum element in remaining unsorted array
        min_index = i
        for j in range(i + 1, n):
            if arr[j] < arr[min_index]:
                min_index = j
        
        # Swap the found minimum element with the first element of unsorted array
        arr[i], arr[min_index] = arr[min_index], arr[i]

# Example usage:
arr = [64, 25, 12, 22, 11]
selection_sort(arr)
print("Sorted array:", arr)
\end{verbatim}

In this implementation: - We iterate through the list and find the index
of the smallest element in the unsorted region. - We swap this smallest
element with the first element of the unsorted region. - We repeat this
process until the entire array is sorted.

Selection Sort is straightforward to implement and understand, making it
suitable for educational purposes or situations where simplicity is
preferred. However, it is less efficient compared to more advanced
sorting algorithms like Merge Sort or Quick Sort, especially for larger
data sets.

\subsection{Bubble Sort}\label{bubble-sort}

Bubble Sort is a simple sorting algorithm that repeatedly steps through
the list, compares adjacent elements, and swaps them if they are in the
wrong order. It passes through the list multiple times until the list is
sorted. Here's an introduction to Bubble Sort:

\subsubsection{Description}\label{description-2}

Bubble Sort gets its name because smaller or larger elements ``bubble''
to the top (beginning) of the list in each pass. It works by repeatedly
comparing adjacent elements and swapping them if they are in the wrong
order. After each pass, the largest (or smallest, depending on the
sorting order) element is guaranteed to be in its correct position. This
process is repeated for each element in the list.

\subsubsection{Steps}\label{steps-2}

\begin{enumerate}
\def\labelenumi{\arabic{enumi}.}
\item
  \textbf{Passes through the List}: Start with the first element and
  compare it with the next element. If they are in the wrong order, swap
  them.
\item
  \textbf{Repeated Passes}: Continue making passes through the list
  until no more swaps are needed, which indicates that the list is
  sorted.
\item
  \textbf{Complexity}: The algorithm's efficiency improves as larger
  elements (or smaller ones) are sorted and ``bubble up'' towards their
  correct positions.
\end{enumerate}

\subsubsection{Example}\label{example-2}

Let's sort the array \texttt{{[}5,\ 2,\ 4,\ 6,\ 1,\ 3{]}} using Bubble
Sort:

\begin{itemize}

\item
  \textbf{Initial Array}: \texttt{{[}5,\ 2,\ 4,\ 6,\ 1,\ 3{]}}
\end{itemize}

\begin{enumerate}
\def\labelenumi{\arabic{enumi}.}

\item
  \textbf{First Pass}: Compare adjacent elements and swap if necessary:

  \begin{itemize}
  
  \item
    \texttt{{[}2,\ 5,\ 4,\ 6,\ 1,\ 3{]}} (swap \texttt{5} and
    \texttt{2})
  \item
    \texttt{{[}2,\ 4,\ 5,\ 6,\ 1,\ 3{]}} (swap \texttt{5} and
    \texttt{4})
  \item
    \texttt{{[}2,\ 4,\ 5,\ 6,\ 1,\ 3{]}} (no swap needed)
  \item
    \texttt{{[}2,\ 4,\ 5,\ 1,\ 6,\ 3{]}} (swap \texttt{6} and
    \texttt{1})
  \item
    \texttt{{[}2,\ 4,\ 5,\ 1,\ 3,\ 6{]}} (swap \texttt{6} and
    \texttt{3})
  \end{itemize}
\item
  \textbf{Second Pass}: Continue with the remaining unsorted elements:

  \begin{itemize}
  
  \item
    \texttt{{[}2,\ 4,\ 1,\ 5,\ 3,\ 6{]}}
  \item
    \texttt{{[}2,\ 4,\ 1,\ 3,\ 5,\ 6{]}}
  \item
    \texttt{{[}2,\ 1,\ 4,\ 3,\ 5,\ 6{]}}
  \item
    \texttt{{[}2,\ 1,\ 3,\ 4,\ 5,\ 6{]}}
  \end{itemize}
\item
  \textbf{Final Pass}: The array is now sorted:

  \begin{itemize}
  
  \item
    \texttt{{[}1,\ 2,\ 3,\ 4,\ 5,\ 6{]}}
  \end{itemize}
\end{enumerate}

\subsubsection{Complexity}\label{complexity-2}

\begin{itemize}
\item\textbf{Time Complexity}: $O(n^2)$ in the worst-case scenario (when the array is reverse sorted) and in the average-case scenario. $O(n)$ in the best-case scenario (when the array is already sorted).
\item\textbf{Space Complexity}: $O(1)$ additional space, as it sorts in-place.
\end{itemize}

\subsubsection{Implementation (Python)}\label{implementation-python-2}

Here's a simple implementation of Bubble Sort in Python:

\begin{verbatim}
def bubble_sort(arr):
    n = len(arr)
    # Traverse through all array elements
    for i in range(n):
        # Last i elements are already in place, so no need to check them
        for j in range(0, n-i-1):
            # Swap if the element found is greater than the next element
            if arr[j] > arr[j+1]:
                arr[j], arr[j+1] = arr[j+1], arr[j]

# Example usage:
arr = [5, 2, 4, 6, 1, 3]
bubble_sort(arr)
print("Sorted array:", arr)
\end{verbatim}

In this implementation: - We iterate over the array multiple times. -
For each pass, we compare adjacent elements and swap them if they are in
the wrong order. - The algorithm terminates early if no swaps are needed
during a pass, indicating that the list is already sorted.

Bubble Sort is straightforward to implement and understand but is
generally inefficient for large data sets compared to more advanced
sorting algorithms like Merge Sort or Quick Sort.

\subsection{Exercises}\label{exercises}

\subsubsection{Insertion Sort Exercises}\label{insertion-sort-exercises}

\begin{enumerate}
\def\labelenumi{\arabic{enumi}.}

\item
  \textbf{Exercise 1: Implement Insertion Sort}

  \begin{itemize}
  
  \item
    \textbf{Problem}: Write a function \texttt{insertion\_sort(arr)}
    that sorts an array \texttt{arr} using Insertion Sort. Use only
    while loops within the function.
  \end{itemize}
\item
  \textbf{Exercise 2: Count Inversions}

  \begin{itemize}
  
  \item
    \textbf{Problem}: Modify the \texttt{insertion\_sort} function to
    count the number of inversions in the array. An inversion is a pair
    of elements \texttt{(arr{[}i{]},\ arr{[}j{]})} such that
    \texttt{i\ \textless{}\ j} and
    \texttt{arr{[}i{]}\ \textgreater{}\ arr{[}j{]}}.
  \end{itemize}
\item
  \textbf{Exercise 3: Sort Characters in a String}

  \begin{itemize}
  
  \item
    \textbf{Problem}: Write a function \texttt{sort\_string(s)} that
    sorts the characters in a string \texttt{s} using Insertion Sort and
    returns the sorted string.
  \end{itemize}
\end{enumerate}

\subsubsection{Bubble Sort Exercises}\label{bubble-sort-exercises}

\begin{enumerate}
\def\labelenumi{\arabic{enumi}.}

\item
  \textbf{Exercise 4: Implement Bubble Sort}

  \begin{itemize}
  
  \item
    \textbf{Problem}: Write a function \texttt{bubble\_sort(arr)} that
    sorts an array \texttt{arr} using Bubble Sort. Can you use only
    while loops?
  \end{itemize}
\item
  \textbf{Exercise 5: Optimized Bubble Sort}

  \begin{itemize}
  
  \item
    \textbf{Problem}: Modify the \texttt{bubble\_sort} function to
    implement an optimized version of Bubble Sort that terminates early
    if no swaps are made in a pass.
  \end{itemize}
\item
  \textbf{Exercise 6: Bubble Sort for List of Lists}

  \begin{itemize}
  
  \item
    \textbf{Problem}: Implement Bubble Sort to sort a list containing
    many lists. Every list (including the outermost one) should be
    sorted.
  \end{itemize}
\end{enumerate}

\clearpage

\section{Sorting Algorithms: Quick, Merge}
\subsubsection{Reflection from Last
Day:}\label{reflection-from-last-day}

\begin{itemize}

\item
  Answer questions about sorting algorithms
\item
  Discuss exercises from previous day
\end{itemize}

Quicksort is a highly efficient sorting algorithm and is based on the
divide-and-conquer strategy. It works by selecting a \texttt{pivot}
element from the array and partitioning the other elements into two
sub-arrays according to whether they are less than or greater than the
pivot. The sub-arrays are then recursively sorted.

\subsubsection{Description}\label{description}

Quicksort follows these steps:

\begin{enumerate}
\def\labelenumi{\arabic{enumi}.}
\item
  \textbf{Choose a Pivot}: Select an element from the array as the
  pivot. There are various ways to choose a pivot, such as picking the
  first element, the last element, a random element, or using a
  median-of-three approach.
\item
  \textbf{Partitioning}: Rearrange the array so that all elements less
  than the pivot are on the left side of the pivot, and all elements
  greater than the pivot are on the right side. After partitioning, the
  pivot is in its final sorted position.
\item
  \textbf{Recursively Apply}: Recursively apply the above steps to the
  sub-arrays of elements with smaller and greater values.
\end{enumerate}

\subsubsection{Steps}\label{steps}

Here's a high-level breakdown of the steps involved in Quicksort:

\begin{itemize}

\item
  \textbf{Step 1}: Choose a pivot element from the array.
\item
  \textbf{Step 2}: Partition the array into two sub-arrays: elements
  less than the pivot and elements greater than the pivot.
\item
  \textbf{Step 3}: Recursively apply Quicksort to the sub-arrays.
\end{itemize}

\subsubsection{Example}\label{example}

Let's sort the array \texttt{{[}10,\ 80,\ 30,\ 90,\ 40,\ 50,\ 70{]}}
using Quicksort:

\begin{itemize}

\item
  \textbf{Initial Array}:
  \texttt{{[}10,\ 80,\ 30,\ 90,\ 40,\ 50,\ 70{]}}
\end{itemize}

\begin{enumerate}
\def\labelenumi{\arabic{enumi}.}
\item
  \textbf{Choose Pivot}: Choose \texttt{50} as the pivot (could be
  chosen randomly).
\item
  \textbf{Partitioning}:

  \begin{itemize}
  
  \item
    After partitioning around the pivot \texttt{50}, the array might
    look like \texttt{{[}10,\ 30,\ 40,\ 50,\ 80,\ 90,\ 70{]}}.
  \item
    Elements less than \texttt{50} are on the left
    (\texttt{{[}10,\ 30,\ 40{]}}), and elements greater than \texttt{50}
    are on the right (\texttt{{[}80,\ 90,\ 70{]}}).
  \end{itemize}
\item
  \textbf{Recursively Sort}:

  \begin{itemize}
  
  \item
    Apply Quicksort recursively to the left sub-array
    (\texttt{{[}10,\ 30,\ 40{]}}) and the right sub-array
    (\texttt{{[}80,\ 90,\ 70{]}}).
  \end{itemize}
\item
  \textbf{Final Sorted Array}: After all recursive calls, the array
  becomes \texttt{{[}10,\ 30,\ 40,\ 50,\ 70,\ 80,\ 90{]}}.
\end{enumerate}

\subsubsection{Complexity}\label{complexity}

\begin{itemize}
\item\textbf{Time Complexity}:
\begin{itemize}
\item $O(n \log n)$ on average, where $n$ is the number of elements.
\item $O(n^2)$ in the worst-case scenario (rare), typically when the pivot selection is poor (e.g., already sorted array).
\end{itemize}
\item\textbf{Space Complexity}: $O(\log n)$ additional space for the recursive call stack.
\end{itemize}

\subsubsection{Implementation (Python)}\label{implementation-python}

Here's a simple implementation of Quicksort in Python:

\begin{verbatim}
def quicksort(arr):
    if len(arr) <= 1:
        return arr
    
    pivot = arr[len(arr) // 2]  # Choosing the middle element as pivot
    left = [x for x in arr if x < pivot]
    middle = [x for x in arr if x == pivot]
    right = [x for x in arr if x > pivot]
    
    return quicksort(left) + middle + quicksort(right)

# Example usage:
arr = [10, 80, 30, 90, 40, 50, 70]
sorted_arr = quicksort(arr)
print("Sorted array:", sorted_arr)
\end{verbatim}

In this implementation: - We recursively partition the array into
smaller sub-arrays based on the pivot element. - Elements less than the
pivot go into the left sub-array, equal elements go into the middle
sub-array, and greater elements go into the right sub-array. - We
concatenate the sorted left, middle, and right sub-arrays to get the
final sorted array.

Quicksort is widely used due to its average-case time complexity of
\(O(n \log n)\), which makes it suitable for sorting large datasets
efficiently. However, care must be taken with its implementation to
avoid worst-case scenarios and ensure optimal performance.

\subsection{MergeSort}\label{mergesort}

Merge Sort is a divide-and-conquer algorithm that divides the input
array into smaller sub-arrays, recursively sorts them, and then merges
the sorted sub-arrays to produce a final sorted array. It is known for
its stable sorting behavior and consistent \(O(n \log n)\) time
complexity.

\subsubsection{Description}\label{description-1}

Merge Sort follows these steps:

\begin{enumerate}
\def\labelenumi{\arabic{enumi}.}
\item
  \textbf{Divide}: Divide the unsorted array into two halves recursively
  until each sub-array contains only one element.
\item
  \textbf{Conquer}: Merge the smaller sorted arrays (sub-arrays) back
  together, ensuring that the merged array remains sorted.
\end{enumerate}

\subsubsection{Steps}\label{steps-1}

Here's a high-level breakdown of how Merge Sort works:

\begin{itemize}

\item
  \textbf{Step 1}: Divide the array into two halves.
\item
  \textbf{Step 2}: Recursively divide each half until each sub-array
  contains only one element (base case).
\item
  \textbf{Step 3}: Merge the sorted sub-arrays back together:

  \begin{itemize}
  
  \item
    Compare the elements from each sub-array and place them in order in
    a temporary array.
  \item
    Continue merging until all elements are sorted and merged into a
    single sorted array.
  \end{itemize}
\end{itemize}

\subsubsection{Example}\label{example-1}

Let's sort the array \texttt{{[}38,\ 27,\ 43,\ 3,\ 9,\ 82,\ 10{]}} using
Merge Sort:

\begin{itemize}

\item
  \textbf{Initial Array}: \texttt{{[}38,\ 27,\ 43,\ 3,\ 9,\ 82,\ 10{]}}
\end{itemize}

\begin{enumerate}
\def\labelenumi{\arabic{enumi}.}

\item
  \textbf{Divide}:

  \begin{itemize}
  
  \item
    Divide the array into two halves: \texttt{{[}38,\ 27,\ 43,\ 3{]}}
    and \texttt{{[}9,\ 82,\ 10{]}}.
  \end{itemize}
\item
  \textbf{Recursively Sort}:

  \begin{itemize}
  
  \item
    Divide further until each sub-array contains only one element:
    \texttt{{[}38{]}}, \texttt{{[}27{]}}, \texttt{{[}43{]}},
    \texttt{{[}3{]}}, \texttt{{[}9{]}}, \texttt{{[}82{]}},
    \texttt{{[}10{]}}.
  \end{itemize}
\item
  \textbf{Merge}:

  \begin{itemize}
  
  \item
    Merge pairs of sub-arrays:

    \begin{itemize}
    
    \item
      Merge \texttt{{[}38{]}} and \texttt{{[}27{]}} to
      \texttt{{[}27,\ 38{]}}.
    \item
      Merge \texttt{{[}43{]}} and \texttt{{[}3{]}} to
      \texttt{{[}3,\ 43{]}}.
    \item
      Merge \texttt{{[}9{]}} and \texttt{{[}82{]}} to
      \texttt{{[}9,\ 82{]}}.
    \item
      Merge \texttt{{[}10{]}} with the already merged
      \texttt{{[}3,\ 43{]}} to \texttt{{[}3,\ 10,\ 43{]}}.
    \end{itemize}
  \item
    Merge the remaining sub-arrays until the entire array is sorted:
    \texttt{{[}3,\ 9,\ 10,\ 27,\ 38,\ 43,\ 82{]}}.
  \end{itemize}
\end{enumerate}

\subsubsection{Complexity}\label{complexity-1}

\begin{itemize}
\item\textbf{Time Complexity}: $O(n \log n)$ in all cases (worst-case, average-case, and best-case scenarios).
\item\textbf{Space Complexity}: $O(n)$ additional space for the temporary array used in merging.
\end{itemize}

\subsubsection{Implementation (Python)}\label{implementation-python-1}

Here's a simple implementation of Merge Sort in Python:

\begin{verbatim}
def merge_sort(arr):
    if len(arr) <= 1:
        return arr
    
    # Divide the array into two halves
    mid = len(arr) // 2
    left_half = arr[:mid]
    right_half = arr[mid:]
    
    # Recursively sort each half
    left_half = merge_sort(left_half)
    right_half = merge_sort(right_half)
    
    # Merge the sorted halves
    return merge(left_half, right_half)

def merge(left, right):
    sorted_arr = []
    i = j = 0
    
    # Compare elements from left and right sub-arrays
    while i < len(left) and j < len(right):
        if left[i] <= right[j]:
            sorted_arr.append(left[i])
            i += 1
        else:
            sorted_arr.append(right[j])
            j += 1
    
    # Append remaining elements
    sorted_arr.extend(left[i:])
    sorted_arr.extend(right[j:])
    
    return sorted_arr

# Example usage:
arr = [38, 27, 43, 3, 9, 82, 10]
sorted_arr = merge_sort(arr)
print("Sorted array:", sorted_arr)
\end{verbatim}

In this implementation: - \texttt{merge\_sort} function recursively
divides the array into halves and sorts each half. - \texttt{merge}
function merges two sorted arrays (\texttt{left} and \texttt{right})
into a single sorted array. - The \texttt{merge\_sort} function calls
\texttt{merge} to combine the sorted halves back together to produce the
final sorted array.

Merge Sort is efficient and stable, making it a popular choice for
sorting large datasets where stability and predictable performance are
important considerations.

\subsection{Exercises}\label{exercises}

\subsubsection{Quick Sort Exercises}\label{quick-sort-exercises}

\begin{enumerate}
\def\labelenumi{\arabic{enumi}.}

\item
  \textbf{Exercise 1: Implement Quick Sort}

  \begin{itemize}
  
  \item
    \textbf{Problem}: Write a function \texttt{quick\_sort(arr)} that
    sorts an array \texttt{arr} using the Quick Sort algorithm. Can you
    implement without the use of recursion?
  \end{itemize}
\item
  \textbf{Exercise 2: Randomized Pivot Selection}

  \begin{itemize}
  
  \item
    \textbf{Problem}: Modify the \texttt{quick\_sort} function to use a
    randomly selected pivot to improve performance on average.
  \end{itemize}
\item
  \textbf{Exercise 3: In-place Quick Sort}

  \begin{itemize}
  
  \item
    \textbf{Problem}: Implement an in-place version of Quick Sort that
    sorts the array without using additional memory for new arrays.
  \end{itemize}
\end{enumerate}

\subsubsection{Merge Sort Exercises}\label{merge-sort-exercises}

\begin{enumerate}
\def\labelenumi{\arabic{enumi}.}

\item
  \textbf{Exercise 4: Implement Merge Sort}

  \begin{itemize}
  
  \item
    \textbf{Problem}: Write a function \texttt{merge\_sort(arr)} that
    sorts an array \texttt{arr} using the Merge Sort algorithm. Can you
    implement without the use of recursion?
  \end{itemize}
\item
  \textbf{Exercise 5: Merge Sort with Uneven Splits}

  \begin{itemize}
  
  \item
    \textbf{Problem}: Implement a bottom-up version of Merge Sort that
    sorts the array iteratively instead of recursively; also the Merge
    Sort should divide the list unevenly during the split operations
    (e.g., 1/4 and 3/4 fraction of the lists).
  \end{itemize}
\item
  \textbf{Exercise 6: Merge Sort using QuickSort}

  \begin{itemize}
  
  \item
    \textbf{Problem}: Implement Merge Sort that uses QuickSort in the
    following way: if the middle item is a balanced pivot choice, then
    use QuickSort. If not, default to Merge Sort.
  \end{itemize}
\end{enumerate}

\clearpage

\section{Review, Recap, Exam}
\subsubsection{Exam/Quiz}\label{examquiz}

The final lecture is a review of the previous lectures. After the
lecture, a final quiz was presented to the students. This exam is purely
for the students to measure their progress and understanding in the
class. However, the student(s) with the highest score received a cash
prize.

\subsubsection{Recap of the Main Topics}\label{recap-of-the-main-topics}

\begin{enumerate}
\def\labelenumi{\arabic{enumi}.}

\item
  \textbf{Types in Python}
\end{enumerate}

Python is a dynamically-typed language, meaning that variables can be
assigned values of any type without specifying the type upfront. The
core data types in Python include integers (\texttt{int}),
floating-point numbers (\texttt{float}), strings (\texttt{str}), and
booleans (\texttt{bool}). Python also supports complex data structures
like lists (\texttt{list}), tuples (\texttt{tuple}), dictionaries
(\texttt{dict}), and sets (\texttt{set}). Understanding these types is
crucial, as they form the building blocks of Python programming. The
flexibility of Python's type system allows for easier and quicker
development, though it requires careful management to avoid type-related
bugs.

\begin{enumerate}
\def\labelenumi{\arabic{enumi}.}
\setcounter{enumi}{1}

\item
  \textbf{For Loops and Recursion}
\end{enumerate}

For loops in Python are used to iterate over sequences like lists,
strings, or ranges. They are essential for performing repetitive tasks
and traversing data structures. Recursion, on the other hand, is a
technique where a function calls itself to solve smaller instances of a
problem. Both are fundamental to algorithm development, though recursion
can be more abstract and sometimes less intuitive than loops.
Understanding when to use loops versus recursion is key to writing
efficient code.

\begin{enumerate}
\def\labelenumi{\arabic{enumi}.}
\setcounter{enumi}{2}

\item
  \textbf{More Methods of Iteration in Python}
\end{enumerate}

Beyond for loops, Python offers several other iteration techniques, such
as while loops, list comprehensions, and generator expressions. List
comprehensions provide a concise way to create lists, while generator
expressions offer memory-efficient ways to iterate over large datasets.
Python's \texttt{itertools} module extends iteration capabilities,
providing tools for creating complex iterators. Mastery of these
iteration methods is essential for writing concise, readable, and
efficient Python code.

\begin{enumerate}
\def\labelenumi{\arabic{enumi}.}
\setcounter{enumi}{3}

\item
  \textbf{Objects, Libraries, Data Science}
\end{enumerate}

Python is an object-oriented language, meaning that everything in Python
is an object. Understanding classes and objects is crucial for
structuring code in a modular and reusable way. Python's extensive
libraries, like NumPy, Pandas, Matplotlib, and SciPy, make it the
language of choice for data science and analytics. These libraries allow
for powerful data manipulation, statistical analysis, and visualization.
Python's ability to integrate seamlessly with data science libraries has
revolutionized fields like machine learning and AI.

\begin{enumerate}
\def\labelenumi{\arabic{enumi}.}
\setcounter{enumi}{4}

\item
  \textbf{Growth of Functions}
\end{enumerate}

In computer science, analyzing the growth of functions is related to
understanding algorithm complexity, often expressed in Big O notation.
This concept helps in evaluating how the performance of an algorithm
scales with the size of the input. Understanding the growth of functions
is critical for writing efficient code, as it directly impacts how well
your programs perform on large datasets.

\begin{enumerate}
\def\labelenumi{\arabic{enumi}.}
\setcounter{enumi}{5}

\item
  \textbf{Searching Algorithms}
\end{enumerate}

Searching algorithms are essential for finding specific elements in data
structures. Common searching techniques include linear search and binary
search. Linear search scans each element sequentially, while binary
search is more efficient, cutting the search space in half with each
step, but it requires the data to be sorted. Understanding these
algorithms is foundational for tasks like database querying, data
processing, and more.

\begin{enumerate}
\def\labelenumi{\arabic{enumi}.}
\setcounter{enumi}{6}

\item
  \textbf{Sorting Algorithms: Bubble, Selection, Insertion}
\end{enumerate}

These three sorting algorithms are fundamental but often considered
inefficient for large datasets. - \textbf{Bubble Sort} repeatedly swaps
adjacent elements that are out of order. - \textbf{Selection Sort}
repeatedly selects the smallest (or largest) element from the unsorted
part and places it in the correct position. - \textbf{Insertion Sort}
builds a sorted array one element at a time by inserting each element
into its proper position.

While these algorithms are easy to understand and implement, they are
generally outperformed by more advanced sorting techniques.

\begin{enumerate}
\def\labelenumi{\arabic{enumi}.}
\setcounter{enumi}{7}

\item
  \textbf{Sorting Algorithms: Quick, Merge}
\end{enumerate}

Quick and Merge Sort are more efficient and commonly used in practice. -
\textbf{Quick Sort} uses a divide-and-conquer approach, selecting a
`pivot' and partitioning the array into elements less than and greater
than the pivot. It's efficient for large datasets but can degrade to
quadratic time in the worst case. - \textbf{Merge Sort} also uses
divide-and-conquer by recursively splitting the array into halves,
sorting them, and then merging them back together. It guarantees a
stable runtime performance, making it reliable for large datasets.
Understanding these algorithms is crucial for performance-critical
applications and is a fundamental part of algorithm study.

\clearpage

\section{Conclusion}

We have covered basic topics about how to use the Python programming language
to reason about and implement basic algorithms on a computer.
\textbf{However, we have only scratched the surface!}
Please refer to the CLRS textbook~\citep{CLRS} for more advanced topics
related to algorithms. Keep learning! Keep pushing!

Please contact the authors about any errors in the lecture notes.
Also, feel free to reach out with suggestions on how to improve the notes.

\section{Acknowledgements}

Daniel Alabi was supported by the Simons Foundation (965342, D.A.) as part of the Junior Fellowship from the Simons Society
of Fellows.
We are also grateful to the following institutions for providing
financial support to NaijaCoder:
IEEE Technical Committee on Data Engineering, the MIT Davis Peace Prize Committee,
the UC Berkeley Mastercard Foundation Scholars program, and the
Mercatus Center at George Mason University.

\bibliographystyle{alpha}
\bibliography{main}

\end{document}